%Paper: hep-ph/9412373
%From: Michael P. Mattis <mattis@skyrmion.lanl.gov>
%Date: Mon, 26 Dec 94 16:18:00 -0700
%Date (revised): Fri, 6 Jan 95 17:58:04 -0700

\input harvmac
%\draftmode
\def\etal{\it et al. \rm}
\def\es{e_{\scriptscriptstyle \rm S}}
\def\bare{{\rm bare}}
\def\Mbare{M_\bare}
\def\Mren{M_\ren}
\def\bra#1{\left\langle #1\right|}
\def\ket#1{\left| #1\right\rangle}

\def\hf{{\textstyle{1\over2}}}
\def\thf{{\textstyle{3\over2}}}
\def\fhf{{\textstyle{5\over2}}}
\def\sst{\scriptscriptstyle}

\def\sqr#1#2{{\vcenter{\vbox{\hrule height.#2pt
        \hbox{\vrule width.#2pt height#1pt \kern#1pt
           \vrule width.#2pt}
        \hrule height.#2pt}}}}
\def\square{\mathchoice\sqr84\sqr84\sqr53\sqr43}

\def\eff{{\rm eff}}
\def\ren{{\rm ren}}
\def\gbare{g_\bare}
\def\gren{g_\ren}
\def\sqn{\sqrt{N_c}}
\def\lam{\Lambda}
\def\laminv{\lam^{-1}}
\lref\Guadagnini{E. Guadagnini, \sl Nucl.~Phys.~\bf B236 \rm (1984) 35.}
\lref\Derrick{G. H. Derrick, \sl J. Math.~Phys.~\bf5 \rm (1964) 1252;
R. Hobart, \sl Proc.~Royal.~Soc. London \bf 82 \rm (1963) 201.}
\lref\LeeTab{See for example
T.-S.~H.~Lee and F.~Tabakin, \sl Nucl.~Phys.~\bf A191 \rm (1972) 332, Table 1;
V.~Barger and M.~Ebel, \sl Phys.~Rev.~\bf138 \rm (1965) B1148;
R.~Machleidt, K.~Holinde and Ch.~Elster, \sl Phys.~Rep.~\bf149 \rm (1987)
1; and M.~Lacombe \etal,  \sl Phys.~Rev.~\bf C21 \rm (1980) 861.}
\lref\Jackson{A. Jackson, A. D. Jackson, and V. Pasquier,
\sl Nucl.~Phys.~\bf A432 \rm (1985) 567.}
\lref\baby{B. Piette, H.. M\"uller-Kirsten, D. Tchrakian and W. Zakrzewski,
\sl Phys.~Lett. \bf B320 \rm (1994) 294;
B. Piette, B. Schroers and W.  Zakrzewski,
hep-th/9406160, to appear in \sl Zeit.~Phys.~\bf C\rm, and  hep-ph/9410256.}
\lref\Othree{For a review and references to the literature
on the $O(3)$ $\sigma$ model, see V. Novikov,
M. Shifman, A. Vainshtein and V. Zakharov, \sl Phys.~Rep.~\bf116 \rm
 (1984) 103.  A solution of the two-dimensional model is given by P. Wiegmann,
\sl Phys.~Lett.~\bf152B \rm (1985) 209.}
\lref\Veneziano{G.~Veneziano, \sl Nucl.~Phys.~\bf B117 \rm (1976) 519.}
\lref\DJMnew{R. Dashen, E. Jenkins and A. V. Manohar,
hep-ph/9411234.}
\lref\Georgi{ C. Carone, H. Georgi and S. Osofsky,
 {\sl Phys.~Lett.} {\bf B322} (1994) 227;
C. Carone, H. Georgi, L. Kaplan
and D. Morin, \sl Phys.~Rev.~\bf D50 \rm (1994) 5793.}
\lref\Manoharold{A. V. Manohar, {\sl Nucl.~Phys.} {\bf B248} (1984) 19.}
\lref\DJM{R. Dashen and A. V. Manohar, {\sl Phys.~Lett.} {\bf B315}
(1993) 425 and 438; R. Dashen, E. Jenkins and A. V. Manohar,
{\sl Phys.~Rev.} \bf D49 \rm (1994) 4713; E. Jenkins and A. V.
Manohar, \sl Phys.~Lett.~\bf B335 \rm (1994) 452.}
\lref\Jenkins{E. Jenkins, {\sl Phys.~Lett.} {\bf B315} (1993) 441.}
\lref\Luty{M. A. Luty and J. March-Russell,  \sl Nucl.~Phys.~\bf B426
\rm (1994) 71;  M. A. Luty, hep-ph/9405271; M. A. Luty, J.
March-Russell and M. White, hep-ph/9405272.}
\lref\tHooft{G. 't Hooft, {\sl Nucl.~Phys.} {\bf B72} (1974) 461 and
{\bf B75} (1974) 461.}
\lref\Syracuse{B. S. Balakrishna, V. Sanyuk, J. Schechter and
A. Subbaraman, \sl Phys.~Rev. \bf D45 \rm (1992) 344; Sec.~II.}
\lref\DiakPet{D. Diakonov, \sl Acta Phys.~Pol.~\bf B25 \rm (1994) 17;
D. Diakonov and V. Petrov, St.~Petersburg preprint LNPI-1394,
1988 (unpublished).}
\lref\Schroers{B. J. Schroers, \sl Zeit.~Phys.~\bf C61 \rm (1994) 479; Sec.~3.}
\lref\ANW{G. Adkins, C. Nappi and E. Witten, {\sl Nucl.~Phys. }
{\bf B228} (1983) 552.}
\lref\MatMuk{M. Mattis and M. Mukerjee, \sl Phys.~Rev.~Lett.~\bf61
\rm (1988)  1344.}
\lref\MatBraat{M. Mattis and E. Braaten, \sl Phys.~Rev.~\bf D39 \rm
(1989) 2737.}
\lref\action{M. Mattis, \sl Phys.~Rev.~ \bf D39 \rm (1989) 994;
\sl Phys.~Rev.~Lett.~\bf 63 \rm (1989) 1455.}
\lref\Skyrme{ T. H. R. Skyrme, {\sl Proc. Roy. Soc.} {\bf A260} (1961) 127;
\sl Nucl.~Phys.~\bf31 \rm (1962) 556.}
\lref\HEHW{A. Hayashi, G. Eckart, G. Holzwarth and H. Walliser,
{\sl Phys.~Lett.} {\bf B147} (1984) 5.}
\lref\vecmesrefs{B. Schwesinger, H. Weigel,
G. Holzwarth and A. Hayashi, \sl Phys.~Rep.~\bf173 \rm (1989) 173; and
references therein.}
\lref\MP{M. Mattis and M. Peskin, {\sl Phys.~Rev.} {\bf D32} (1985) 58;
M.~Karliner, \sl Phys.~Rev.~Lett.~\bf57 \rm (1986) 523.}
\lref\ninej{M. Mattis, \sl Phys.~Rev.~Lett.~\bf 56 \rm (1986) 1103.}
\lref\GerSak{ J. Gervais and B. Sakita, {\sl Phys.~Rev.} {\bf D30}
(1984) 1795.}
\lref\DashMan{R. Dashen and A. Manohar,
{\sl Phys.~Lett.} {\bf B315} (1993) 425.}
\lref\Schulman{ L. S. Schulman, \sl Phys.~Rev. \bf 176 \rm (1968) 1558; and
 {\sl Techniques and Applications of Path
Integration}, (Wiley-Interscience 1981). }
\lref\Coleman{ S. Coleman, {\sl Phys.~Rev.} {\bf D11} (1975) 2088;
S. Mandelstam, \sl Phys.~Rev.~\bf D11 \rm (1975) 3026.}
\lref\DHM{N. Dorey, J. Hughes and M. Mattis, \sl Phys.~Rev.~\bf D50
\rm (1994) 5816.}
\lref\KogWil{A classic review is J. Kogut and K. Wilson, \sl
Phys.~Rep.~\bf 12\rm, No.~2 (1974) 75.}
\lref\Chodos{A. Chodos and C. Thorn, {\sl Phys.~Rev.} {\bf D12} (1975) 2733.}
\lref\RGB{M. Rho, A. S. Goldhaber
and G. E. Brown, {\sl Phys.~Rev.~Lett.} {\bf 51} (1983) 747.}
\lref\GoldJaf{J. Goldstone and R. L. Jaffe,
 {\sl Phys.~Rev.~Lett.} {\bf 51} (1983) 1518.}
\lref\Rho{An excellent recent review of chiral bags and Cheshire
cats is M. Rho, \sl Phys.~Rep.~\bf 240 \rm (1994) 1.}
\lref\Nielsen{S. Nadkarni, H. Nielsen, and I. Zahed,
\sl Nucl.~Phys.~\bf B253 \rm (1985) 308;
P.H. Damgaard, H.B. Nielsen and R. Sollacher, \sl Nucl.~Phys.~\bf B385
\rm (1992) 227 and \sl Phys.~Lett.~\bf B296 \rm (1992) 132.}
\lref\DHMPRL{N. Dorey, J. Hughes and M. Mattis, \sl Phys.~Rev.~Lett. \bf
73 \rm (1994) 1211.}
\lref\ArnMat{P. Arnold and M. Mattis, \sl Phys.~Rev.~Lett.~\bf65 \rm
(1990)  831; M. Mattis and R. Silbar, \sl Phys.~Rev.~\bf D \rm (in press).}
\lref\ManPLB{A. Manohar,  \sl Phys.~Lett.~\bf B336 \rm (1994) 502.}
\lref\Witten{ E. Witten, \sl Nucl.~Phys.~\bf B160 \rm
(1979) 57.}
\font\authorfont=cmcsc10 \ifx\answ\bigans\else scaled\magstep1\fi
%\divide\baselineskip by 3
%\multiply\baselineskip by 4
%\advance\topskip by -1in\relax
{\divide\baselineskip by 3
\multiply\baselineskip by 2
\def\prenomat{\matrix{\rm hep-ph/9412373&\cr
\rm SWAT/60&\cr}}
\Title{$\prenomat$}{\vbox{\centerline{From Effective Lagrangians,
to Chiral Bags,  to Skyrmions}\vskip2pt
\centerline{with the Large-$N_c$ Renormalization Group}}}
\centerline{\authorfont Nicholas Dorey}
\bigskip
\centerline{\sl Physics Department, University College of Swansea}
\centerline{\sl Swansea SA2$\,$8PP UK $\quad$ \tt pydorey@swansea.ac.uk}
\bigskip
\centerline{and}
\bigskip
\centerline{\authorfont Michael P. Mattis}
\bigskip
\centerline{\sl Theoretical Division T-8, Los Alamos National Laboratory}
\centerline{\sl Los Alamos, NM 87545 USA$\quad$ \tt mattis@skyrmion.lanl.gov}
\vskip .3in
\noindent
We explicitly relate
effective meson-baryon Lagrangian models, chiral bags, and Skyrmions
in the following way. First, effective Lagrangians are constructed in a manner
consistent with an underlying large-$N_c$ QCD. An infinite
set of graphs dress the bare Yukawa couplings at
\it leading \rm order in $1/N_c,$ and are summed using
semiclassical techniques. What emerges is
 a picture of the large-$N_c$ baryon reminiscent of the
chiral bag: hedgehog pions for $r\ge\laminv$ patched onto
bare nucleon degrees of freedom for $r\le\laminv,$ where the ``bag radius''
$\laminv$ is the UV cutoff on the graphs.
Next, a novel renormalization group (RG) is derived,
in which the bare Yukawa couplings, baryon masses and hyperfine baryon
mass splittings run with $\lam.$ Finally, this RG flow is shown to
act as a \it filter \rm on the renormalized Lagrangian
parameters: when they are fine-tuned to obey Skyrme-model relations
the continuum limit $\lam\rightarrow\infty$
exists and is, in fact, a Skyrme model; otherwise there is no
continuum limit.
%\looseness=-1
%\relax
\vskip .1in
\Date{\bf 26 December 1994} %replace this line
%by \draft  for preliminary versions or specify \draftmode at some point
\vfil\break
}
\newsec{Introduction}
\subsec{Effective hadron Lagrangians {\rm versus} Skyrmions}
In the absence of reliable quantitative methods for computing the
low-energy properties of QCD, a wide variety of phenomenological models
of the nucleon have emerged and flourished. One general class
of models, which predates QCD by some 30 years, starts from an
\it effective Lagrangian \rm for the baryon and meson fields. The
hope in this approach is that the relevant physics is contained
in the complete set of hadron Feynman diagrams. An orthogonal approach,
pioneered by Skyrme in the early 1960s, uses topology: the baryon is
viewed as a soliton, or \it Skyrmion\rm, in the field of
mesons \refs{\Skyrme,\ANW}.

On the face of it, the two approaches could not be more opposite.
Baryon number in an effective quantum field theory
of hadrons is simply the Noether charge associated with a $U(1)$
symmetry of the Lagrangian. In contrast, in the Skyrmion picture,
baryon number is not associated with any continuous symmetry but
is  instead a topological invariant: the \it winding number \rm
of a meson field configuration. And the physics of Skyrmions
is expressed, not in the language of Feynman diagrams,
but in vocabulary more appropriate to solitons and other extended objects:
collective coordinate quantization, symmetry classification of
small fluctuations about the soliton, and so forth.

In this paper, we exhibit a precise connection between these
two disparate approaches (see Fig.~1).
The bridge between them is  built by combining in a new
 way two important theoretical constructs: 't Hooft's
$1/N_c$ expansion \tHooft, $N_c$ being
the number of colors in the underlying gauge theory, and Wilson's
renormalization group \KogWil. The parameter
 governing the flow of this so-called {\it large-$N_c$
renormalization group} \refs{\DHMPRL\DHM}  is an ultraviolet
cutoff $\Lambda\sim N_c^0$ which regulates the divergences in the effective
Feynman diagrams. Our main result is that soliton models and
(suitably fine-tuned) effective Lagrangian models
are \it completely equivalent \rm
at leading order,\foot{From now on, ``leading order''
refers to the $1/N_c$ expansion. The need to fine-tune the renormalized
Lagrangian couplings is explained in Sec.~1.4 below.}
 in the ``continuum limit'' $\lam\rightarrow
\infty$ (to borrow lattice terminology). In this Section we introduce the
key ideas behind the large-$N_c$ renormalization group;
the remainder of the paper is primarily devoted to exhibiting its solutions
in a series of simple models.

\subsec{The large-$N_c$ limit and semiclassical physics}
In 't Hooft's original formulation, the $1/N_c$ expansion is studied
directly in QCD; as $N_c\rightarrow
\infty$ the physics is dominated by the quenched planar
quark-gluon graphs.
But the large-$N_c$ limit is actually much more predictive when
the lessons of planar QCD are implemented,
not at the fundamental quark-gluon level, but rather at the level of
phenomenological models such as above. Thus, on the one hand,
in effective hadron Lagrangians, it implies a set of \it large-$N_c$
selection rules \rm which place powerful constraints on the allowed particle
spectrum and couplings contained in one's model. As  examples of such
rules \refs{\Veneziano,\Witten},
  whereas meson masses typically scale like $N_c^0$, baryons
(being made up of $N_c$ quarks) have masses that grow linearly with $N_c$.
And whereas Yukawa couplings are strong, growing like $\sqrt{N_c},$
meson self-interactions are weak, with $n$-meson vertices disappearing
like $N_c^{1-n/2}.$ The complete set of these large-$N_c$ selection
rules is reviewed at the beginning of Sec.~2 below.
On the other hand, in Skyrmion physics, $1/N_c$ plays an ostensibly
different role, which is more easily summarized: it parametrizes the
\it semiclassical \rm expansion about the soliton. This is because
$1/N_c$ always enters into Skyrmion Lagrangians in the combination
$\hbar/N_c.$

In fact, this difference is illusory:
 the effective Lagrangian approach, too, becomes semiclassical
in the large-$N_c$ limit (a key to the equivalence of the two pictures).
 What we mean by this is twofold.
First, in calculating the leading-order
contribution to meson-baryon
Green's functions, the naive, graph-by-graph, perturbative
 method fails, and one is
forced instead to sum an infinite class of diagrams. Second, this
sum may be accomplished by solving  \it classical \rm equations
of motion for the meson fields in the background of the baryon source.

In order that these two points be understood,
let us be very explicit at this stage,
 and in so doing, introduce the central field-theoretic problem
of this paper. Consider the bare Yukawa coupling  $\gbare$
depicted in Fig.~2a, which scales like $\sqn$ as stated above.
In principle, we would like to sum all radiative corrections to
this vertex such as Figs.~2b-d, and thereby extract the renormalized
Yukawa coupling $\gren,$ to leading order in $1/N_c.$\foot{The
calculation of  renormalized single-meson emission/absorption from
the nucleon is the simplest arena for our semiclassical methods,
which may also be applied to more complicated processes such as
meson-baryon and baryon-baryon scattering \refs{\DHMPRL,\ArnMat}.}
Focus first on Fig.~2b. It contains two factors of $\gbare,$ therefore two
factors of $\sqn,$ as well as a 3-meson vertex which goes like $1/\sqn$;
so this graph too scales like $\sqn,$ and is a
{\it leading-order}  correction to $\gbare.$ Likewise
Fig.~2c scales like $\sqn$, as the reader can check by multiplying
all vertices together.  An example of a subleading
correction is Fig.~2d. Unlike the others it contains a purely
mesonic loop (indicated by the arrow), hence two extra factors of $1/\sqn$
uncompensated by extra Yukawa vertices. So Fig.~2d and the like are
$1/N_c$ corrections, and will not concern us further.

A moment's thought
confirms the rule: the leading-order dressings are the \it infinite
set \rm of diagrams which contain
only meson \it trees \rm if the baryon line is erased. And the sum of all
trees---like a soliton---is given by the solution to a classical
equation of motion, which we write down in Sec.~3 below.
 The role of the bare large-$N_c$  baryon in this equation is that of
a heavy, slow-moving source, smeared out over a length-scale $\laminv$
so as to cut off the short-distance divergences in the
original Feynman graphs.  As we shall see, the solution of the
regulated equation is a hedgehog cloud of pions and/or other
mesons for $r\ge\laminv,$
glued onto the bare nucleon degrees of freedom which are
restricted to $r\le\laminv.$ The
energy of this cloud renormalizes the mass of the baryon ($\Mbare\rightarrow
\Mren$)
while its large-distance behavior determines the physical
Yukawa coupling ($\gbare\rightarrow\gren$).

\subsec{Of chiral bags and Cheshire cats}
The resulting picture of the
meson-dressed large-$N_c$ baryon is highly reminiscent of
yet a third class of phenomenological models, the
 {\it chiral bag models} \refs{\Chodos\RGB\GoldJaf{--}\Rho}.
 These, too, are hybrid
descriptions of the dressed baryon, in which explicit quark (rather
than nucleon) degrees of freedom
inside a bag of radius $R$ are matched onto an effective theory of hedgehog
pions outside the bag. Even this presumably important distinction
between `nucleon' versus `quark' degrees of freedom inside the bag
disappears as $N_c\rightarrow\infty.$\foot{This observation was originally
made by Witten (Ref.~\Witten, Secs.~5 and 9), and exploited by Gervais and
Sakita (Ref.~\GerSak, Sec.~V). These two papers are highly recommended
background reading, as they too are concerned primarily with
 the semiclassical nature of large-$N_c$. In particular Gervais and
Sakita were the first to study chiral-bag-type structures in this
limit, although not from our effective hadron Lagrangian starting point.}
 For, in this limit, the $N_c$
quarks may be treated in Hartree approximation, and their individual
wave functions effectively
condense into a common mean-field wave function, which we
may identify with the ``wave function of the nucleon.'' Outside
the bag, the analogy is closer still: the pion field
configuration is again determined by solving a nonlinear field equation
coupled to a static source at $r=R$.  The only significant difference
between our composite meson-dressed large-$N_c$ baryon and the traditional
chiral bag is this: our composite baryon follows solely from large $N_c$
and has \it nothing whatsoever
to do with chiral symmetry! \rm For this reason we shall refer to
it as a ``chiral bag'' (Fig.~1 again),
 being careful to retain the quotation marks to
avoid confusion with the traditional chiral bag.

Associated with the traditional chiral bag in the recent literature is the
so-called {\it Cheshire cat principle} \refs{\Nielsen,\Rho}, which
 states that physical quantities
should be independent of the size and shape of the bag. This proposal
is motivated by the success of bosonization in 1+1 dimensions, which
provides an exact mapping between, for instance, the elementary fermion of the
massive Thirring model and the soliton of the sine-Gordon
equation \Coleman.  In its most extreme form, it implies that $R$ can
safely be set to zero, yielding a description of the
nucleon purely in terms of pion fields. One may  conjecture that
this limit is nothing other than the Skyrme model \refs{\GoldJaf,\Rho}.
However, since no equivalent of bosonization has been made to work in
3+1 dimensions, the theoretical status of the Cheshire cat principle
has remained uncertain.

{}From our present large-$N_c$ perspective, the bag
radius $R$ is best reinterpreted as the short-distance cutoff $\laminv$ of the
large-$N_c$ effective theory.
It is then easy to see the significance of the
Cheshire cat principle: the cutoff-independence of physical quantities
is usually referred to as renormalization group
invariance \KogWil. However, except in special cases,
renormalization group invariance of the physical masses and couplings
never comes for free;
the corresponding bare quantities must be varied simultaneously with the
cutoff. As stated earlier, we will refer to this program as the
{\it large-$N_c$
renormalization group}, and devote considerable space to mapping out
its solutions. To our understanding, the concept of such a flow
is not only absent from the usual Cheshire cat philosophy, but in fact
orthogonal to it; $R$ in that scheme is nothing more than a
gauge-fixing parameter.

\subsec{The large-$N_c$ renormalization group as a ``filter''}

The limit of zero bag size, $R\rightarrow 0$ or $\lam\rightarrow\infty$,
corresponds to removing the cutoff completely. This
continuum limit, if it can be taken at all, is by definition
 a UV stable fixed point of the renormalization group
flow.\foot{Because we are implementing renormalization group invariance
strictly at leading order in $1/N_c,$
 the limit $\lam\rightarrow\infty$ is to be taken
\it after \rm $N_c\rightarrow\infty$. We suspect that these limits
do not commute.}
 In  light of the above discussion, it is natural to
conjecture that such a fixed point exists if and only if
the homogeneous meson field
equations ($i.e.,$ with the baryonic source set to zero)
supports a soliton solution.\foot{The only exception we have found
to this rule is a theory of non-self-interacting mesons (Sec.~4).}
 Thus we propose that
variants of the Skyrme model describe
the possible continuum limits of effective Lagrangian theories of mesons and
baryons in the large-$N_c$ limit.

On reflection, there is an obvious counting problem with this
scenario:  effective Lagrangians always contain more free parameters
than the corresponding soliton models. 	Thus, in the former, the physical
masses and couplings are all independent, while in the
latter, there exist non-trivial relations among them; for instance,
the Yukawa constant $\gren$  is completely determined by the
meson self-couplings \ANW. (This feature, of course, is \it precisely the
point \rm of the Skyrmion approach.) If our proposal is correct, it
follows that unless the renormalized parameters in the effective
Lagrangian are tuned
{\it exactly} to those of the corresponding soliton model, there must be
some mathematical obstacle to taking the continuum limit. In other words,
we conjecture that the
large-$N_c$ renormalization group  acts as a {\it filter},
blocking the path to the continuum except for a measure-zero
 subset of the space of renormalized parameters. This filter idea
is the central theme of this paper, and is explicitly realized in
the models to follow. The attentive reader will recognize in the
phrasing of this counting problem the classic symptom of the existence
of an ``irrelevant operator'' \KogWil; naturally this operator turns
out to be the bare Yukawa coupling itself.

\subsec{The plan of this paper}

The paper is organized as follows. Sections 2 and 3
recapitulate our recent Letter \DHMPRL, in somewhat more detail, and
at a more leisurely pace. (Independently, Manohar \ManPLB\
has reached similar conclusions to Ref.~\DHMPRL; other relevant
precursors are Refs.~\ArnMat\ and \GerSak.) In particular, Sec.~2
reviews the large-$N_c$ selection rules mentioned above, and gives
the recipe for constructing large-$N_c$-compatible effective
Lagrangians. Section 3 is devoted to semiclassics: the problem
posed in Fig.~2 is solved completely at a formal level, the issue
of meson contributions to the baryon self-energy is examined,
the hedgehog structure of the meson cloud is revealed,
and the large-$N_c$ renormalization group is defined.

\def\Lmeson{L_{\rm meson}}
\def\calLmeson{{\cal L}_{\rm meson}}
\def\crit{{\rm crit}}
\def\lcrit{\lam_\crit}
\def\fpi{f_\pi}
The reader already familiar with the contents of Ref.~\DHMPRL\ is encouraged
to skip directly to Secs.~4-8, in which the large-$N_c$ renormalization
group is applied to a series of effective hadron models. What
distinguishes these models from one another is our choice of the purely
mesonic piece of the action, $\calLmeson.$  In Sec.~4, $\calLmeson$
is simply the free pion Lagrangian. Not surprisingly, one finds
essentially no running of the bare $\pi N$ Yukawa coupling: $\gbare(\lam)
\cong\gren.$
Much more interesting is the model of Sec.~5, in which
$\calLmeson\,=\,{\fpi^2\over16}\,\Tr\,\partial_\mu U^\dagger
\partial^\mu U\,$, the leading term in chiral perturbation theory.
This is a clean initial test of the ``filter'' idea conjectured
above, because this model is known \it not \rm to support a soliton,
thanks to Derrick's theorem \Derrick. And indeed, we discover a critical
value of the cutoff, $\lcrit\cong 340\,$MeV, beyond which the large-$N_c$
renormalization group cannot be pushed. In Sec.~6 we construct
an \it analytically soluble \rm 2+1 dimensional  model, in
which the would-be Skyrmion is simply the \it instanton \rm of the $O(3)$
$\sigma$ model
in one lower dimension. In this model the physics is all that one might
have hoped: unlike Sec.~5
there is no ultraviolet obstruction to the large-$N_c$
renormalization group; the bare coupling $\gbare(\lam)$ is irrelevant,
flowing to
zero like $\lam^{-4}$; and a ``toy'' Skyrme model indeed lies at the
end of the large-$N_c$ renormalization group trajectory, at which
point explicit baryon number  effectively transmutes into the winding number
of the meson cloud.

Finally, in Secs.~7-8 we augment the nonlinear
pion Lagrangian of Sec.~5 with the 4-derivative ``Skyrme term'' and
study the large-$N_c$ renormalization group, both in the measure-zero
case when the physical Yukawa coupling is tuned precisely to its Skyrme-model
value (Sec.~7), and in the generic case when they differ (Sec.~8).
In a surprising way, involving essential singularities in Skyrme's
equation, and \it local instabilities \rm that develop
in the pion cloud for $\lam\ge\lcrit$ (see the Appendix for
technical details), our ``filter'' conjecture is borne out.

In light of this already long Introduction, we  will spare the
reader a Conclusions section, and commend him instead to keep the
``large-$N_c$ renormalization group as filter'' idea firmly in mind
as he works his way through the examples.

\newsec{Constructing large-$N_c$ effective Lagrangians}
\def\cl{{\rm cl}}
\def\Leff{L_\eff}
\def\Lbaryon{L_{\rm baryon}}
\def\Lyukawa{L_{\rm yukawa}}
\def\Lseagull{L_{\rm seagull}}
\def\calLskyrme{{\cal L}_{\rm skyrme}}
\def\Ibare{{\cal I}_\bare}
\def\Mcl{M_\cl}
\def\Icl{{\cal I}_\cl}
\def\Iren{{\cal I}_\ren}
\subsec{Large-$N_c$ selection rules}
 As mentioned in Sec.~1, the large-$N_c$ limit imposes several
stringent requirements on the allowed spectrum and interactions of
hadrons, which we now review. These rules can be derived quite
independently from several different approaches: not just a
planar-diagrammatic analysis of
large-$N_c$ QCD \refs{\Veneziano,\Witten,\Luty,\Georgi}
 but also the Hartree approximation originally employed
by Witten \Witten, the Skyrme model \refs{\ANW,\GerSak,\MatMuk},
 the non-relativistic quark model \refs{\Manoharold\MatBraat{--}\DJMnew},
 and finally the self-consistency of the
effective hadron Lagrangian \refs{\GerSak,\action}
 as recently emphasized in a series
of interesting papers by Dashen, Jenkins and Manohar \refs{\DJM,\Jenkins}.
The following rules (the first two of which were already invoked
in Sec.~1) should be considered robust,
model-independent features of the large-$N_c$ limit:

\bf(i) \rm
As noted originally by Veneziano, purely mesonic vertices with $n$ external
 legs scale like $N_c^{1-n/2}$ \refs{\Veneziano,\Witten}.
 An important example with $n=1$
is that the pion decay constant $f_{\pi}\sim \sqrt{N_c}$, whereas
meson masses ($n=2$) generically scale like $N_c^0.$

\bf(ii) \rm
Vertices with two baryon legs and $n$ meson legs also scale like
$N_c^{1-n/2}$ so that baryon masses ($n=0$) and Yukawa couplings
($n=1$) grow like $N_c$ and $\sqrt{N_c}$, respectively \refs{\Witten,
\GerSak,\Luty,\Georgi}.

 \bf(iii) \rm
The 2-flavor stable baryon spectrum of large-$N_c$ QCD (with $N_c$ odd)
consists of
an infinite tower of positive parity states with $I=J=\hf$, $\thf$,
$\fhf,\cdots$.\foot{In the Skyrme-model representation this tower
is truly infinite whereas in the quark-model representation it tops
off at $N_c/2$; see Ref.~\DJMnew\ for a detailed discussion of how
to translate Skyrme-model operators into quark-model operators and
\it vice versa \rm in light of this difference.}  To leading order
these states are degenerate \refs{\ANW,\GerSak,\Luty,\Georgi,\DJM},
 with bare mass $M_{\rm bare}\sim N_c$
(as baryons are made of $N_c$ quarks). Hyperfine baryon mass splittings
have the form $J(J+1)/2{\cal I}_{\rm bare}$ where ${\cal I}_{\rm bare}
\sim N_c$ \refs{\ANW,\Luty,\Georgi,\Jenkins}.
 In the Skyrme model these states correspond to (iso)rotational
excitations of the static hedgehog and the splitting term corresponds
exactly to the rotational kinetic energy of the Skyrmion \ANW.

 \bf(iv) \rm
In contrast, the spatial extent of a baryon does not grow but has
a smooth $N_c$-independent limit as $N_c\rightarrow\infty$ as do
the various baryon form factors such as electromagnetic charge distributions,
and baryon number density itself \refs{\Witten,\ANW}.

 \bf(v) \rm
Yukawa couplings are constrained to obey the ``proportionality rule'' which
fixes the interaction strength of
a given meson with each member of the baryon tower as a multiple of one
overall coupling constant \refs{\ANW,\GerSak,\MatMuk,\MatBraat,\DJM}
 (e.g., $g_{\pi NN}\propto g_{\pi N\Delta}\propto g_{\pi\Delta \Delta}
\cdots\,$), up to corrections of order $1/N_c^2$ \DJM.

 \bf(vi) \rm
Finally, the allowed couplings of mesons to the baryon tower
must obey the $I_t=J_t$ rule \refs{\MatMuk,\MatBraat,\action}. For example
the $\rho$ meson must be tensor-coupled to the nucleon
while the $\omega$ meson is vector-coupled at leading order in
$1/N_c$ \refs{\action,\Georgi,\DJMnew}, in good agreement with
phenomenology \LeeTab.  When crossed from the $t$-channel to
the $s$-channel, this rule also implies nontrivial model-independent
relations among meson-baryon $S$ matrix elements \refs{\HEHW\MP{--}\ninej},
 which are tested against the experimental data
 in Refs.~\refs{\MP{--}\ninej}.\foot{Rules \bf (v) \rm
and \bf (vi) \rm are elementary examples of ``large-$N_c$ group theory,''
the state-of-the-art phenomenological predictions of which are summarized
in Ref.~\DJMnew. The existence of such group-theoretic relations may be
traced to the fact that $SU(2N_F)$ spin$\times$flavor symmetry becomes
exact as $N_c\rightarrow\infty$ \refs{\GerSak,\DJM}. This
 is one of the two main attractions of  large-$N_c$
physics, the other being its amenability to a semiclassical treatment.}

These selection rules must be implemented in any large-$N_c$-compatible
effective hadron Lagrangian. What we stressed in our Letter
\DHMPRL\ and will review below is the \it consistency \rm of these rules,
meaning that once they are incorporated into the bare Lagrangian they
continue to hold for physical, renormalized quantities as well.
Thus, while the baryon mass spectrum is renormalized at leading order
by the interactions with the mesons, the $J(J+1)$ structure discussed
in \bf (iii) \rm is preserved, and so, too, the form of the
${\cal O}(\sqn)$ Yukawa couplings of the mesons to the baryon tower
dictated by \bf (v) \rm and \bf (vi)\rm.  And while the ``bare''
nucleon size (as given by the naive ultraviolet cutoff $\laminv\sim N_c^0$)
is effectively enlarged by the meson cloud, it remains of order $N_c^0$
as per \bf (iv)\rm.
This is because the spatial extent of the cloud is dictated by the parameters
\bf (i) \rm of the meson Lagrangian, for instance $m_\rho^{-1}$, or
the product of Skyrme-model parameters $(\es\fpi)^{-1}$ as we shall see
in Secs.~7-8 below.

In general, an effective meson-baryon  Lagrangian is a sum of four parts:
\eqn\Leffdef{\Leff\ =\ \Lbaryon+\Lmeson+\Lyukawa+\Lseagull\ .}
In light of the above selection rules,
let us discuss, in turn, the proper construction of each of these parts.

\subsec{Constructing $\Lbaryon.$} We start with a relativistic baryon
Lagrangian for the tower described in \bf (iii) \rm above:
\eqn\Lbaryonrel{\bar N(i\gamma^\mu\partial_\mu-M_N)N
\ +\ \hbox{(higher }I=J\hbox{ baryons)}}
where $N$ means $\left({p\atop n}\right)$. We will need to recast this
unwieldy infinite sum in a more useful form. Since large-$N_c$ baryons are
heavy, it
is natural to split their propagators in the usual way into forwards-in-time
($U$-spinor) plus backwards-in-time ($V$-spinor)
pieces. The latter account for the so-called $Z$-graph contributions
to Feynman diagrams, which turn out to be subleading as we review momentarily
(Sec.~2.5). The remaining time-ordered diagrammatics is one in which
$Z$-graphs have been eliminated, and with them the higher components
of the baryon-antibaryon Fock space. In this way, baryon \it quantum
 field theory \rm collapses to baryon \it quantum  mechanics\rm.
But we can simplify the physics even further. For processes involving
one baryon only, interacting with an arbitrary number of mesons, it is
natural to work in or near the baryon's rest frame, in which case, finally,
\def\bX{{\bf X}}
\eqn\lbaryondef{\Lbaryon\ =\
-\Mbare+\hf\Mbare\dot\bX^2+\Ibare\Tr \dot A^\dagger\dot A+\cdots,}
the dots denoting $1/N_c$ corrections. The first two terms on the right-hand
side are the usual nonrelativistic approximation to the relativistic
mass-energy, $\bX(t)$ being the baryon's position.

The third term, the (iso)rotational kinetic energy,
 is perhaps less familiar. It represents free motion on
the baryon's spin/isospin manifold (just as the previous term denotes
free spatial translation), with $A(t)\in SU(2)$ being the baryon's spin/isospin
collective coordinate. The full meaning of this term is revealed in the
beautiful path integral identity due to Schulman \Schulman:
\def\calD{{\cal D}}
\eqn\schulmanid{\eqalign{&\int_{A(t_1)=A_1}^{A(t_2)=A_2}\calD A(t)\,\exp
i\int dt\big(-\Mbare+\Ibare\Tr\dot A^\dagger\dot A\big)\cr& =\
\sum_{J={1\over2},{3\over2},\cdots}\ \sum_{i_z,s_z=-J}^J
{\textstyle\big\langle A_2\big|{I=J\atop i_z\,s_z}\big\rangle}
\,e^{-i(t_2-t_1)\Mbare^J}\,
{\textstyle\big\langle {I=J\atop i_z\,s_z}\big|A_1\big\rangle}
\ ,}}
where
\eqn\MbareJdef{\Mbare^J\ =\ \Mbare+{J(J+1)\over2\Ibare}\ .}
In other words, $\Ibare\Tr\dot A^\dagger\dot A$ is convenient shorthand for the
free propagation of an \it infinite tower \rm of $I=J$ baryons with
the mass spectrum of a rigid rotor---exactly as required by rule
\bf (iii) \rm given above.  We will adopt Skyrme-model
nomenclature and refer to $\Ibare$ as the ``bare moment of inertia''
of the baryon \ANW. The brackets
\eqn\bracketdef{
{\textstyle\big\langle A\big|{I=J\atop i_z\,s_z}\big\rangle}\ =\
(2J+1)^{1/2}(-)^{J-s_z}D^{(J)}_{-s_z,i_z}(A^\dagger)}
in Eq.~\schulmanid\
are just the change-of-basis overlaps between the usual spin-isospin
baryon representation (the nucleons with $I=J=\hf,$ the $\Delta$'s with
$I=J=\thf,$ etc., with spin and isospin $z$ components $s_z$ and $i_z$),
 and baryon states $\ket{A}$ sharp instead in the
spin/isospin collective coordinate.
The $\ket{A}$ basis too, while popularized by the Skyrme model \ANW,
is useful more generally in large-$N_c$ physics \refs{\GerSak,\Manoharold,
\MatBraat}.
\def\fpinv{\fpi^{-1}}

\subsec{Constructing $\Lmeson$.}
As for the meson piece of the action, it is best to leave $\Lmeson$
completely unspecified for the time being, subject only to the
scaling rule \bf (i) \rm discussed above. Importantly, if one then
rescales all meson fields by a mass parameter proportional to
$\sqn$ (such as the pion decay constant $\fpi$),
a multiplicative factor of $N_c/\hbar$ sits in front of the meson action.
So a leading-order analysis in the $1/N_c$ expansion is tantamount to a
\it semiclassical \rm ($\hbar\rightarrow0$) treatment of the mesonic part
of the path integral. We will exploit this feature shortly.
\def\NN{{\sst NN}}
\def\gpnn{g_{\pi\NN}}
\def\gpnnbare{\gpnn^\bare}

\subsec{Constructing $\Lyukawa$.}
Next, consider $\Lyukawa.$
For large $N_c$, the usual pseudovector coupling of the
pion to the nucleon,\foot{We have absorbed the normally explicit factor
of $(2M_N)^{-1}$ into the pseudovector coupling constant $\gpnnbare,$
which therefore has dimensions of length.
The reason that pseudovector coupling is far preferable to pseudoscalar
coupling in large-$N_c$ physics is that the latter involves an
awkward cancellation: $\gamma^5$ couples the `large' to the `small'
components of the nucleon spinor, the `small' components being
down by $1/N_c,$ which compensates the fact that the
pseudo\it scalar \rm constant grows like $N_c^{3/2}$ as dictated
by the Goldberger-Treiman relation.}
\def\vpi{\vec\pi}
\def\vtau{\vec\tau}
\def\vrho{\vec\rho}
\eqn\PVdef{\gpnnbare\,\partial_\mu\vpi\cdot\bar N\gamma^\mu
\gamma^5\vtau N\ ,}
must be augmented by similar couplings to the entire $I=J$ tower of baryons
in a manner fixed by the proportionality rule
\bf (v)\rm. As before, the $\ket{A}$
basis for the baryons permits an especially compact representation
of this set of couplings, namely \refs{\ANW,\DHM}
\def\bx{{\bf x}}
\eqn\PVAdef{3\gpnnbare\,{\partial\over\partial {x^l}'}\pi^a(x)\,
D^{(1)}_{al}\big(\hat A(t')\big)\,\delta^3(\bx')\ .}
Here the primed space-time coordinate $x'=(t',\bx')$ is the
Poincar$\acute{\rm e}$ transformation of $x$ into the center-of-mass
frame of the baryon, assumed to be moving with fixed velocity
$\dot\bX$ relative to the Lab frame;
 when $\dot\bX\ll1$ so that Lorentz contractions are irrelevant,
one simply has
\eqn\primedef{\bx'\ \approx\ \bx-\bX(t)\ .}
The rotation matrix
$D^{(1)}(\hat A)$ is an operator (hence the `hat' on the $A$) on the
spin and isospin quantum numbers of the single-baryon Hilbert space.
It is completely specified by its
 matrix elements in the conventional spin-isospin baryon basis:
\eqn\Ddef{\eqalign{
{\textstyle\big\langle{I'=J'\atop i'_z\,s'_z}}
\big|D^{(1)}_{al}(\hat A)\big|
{\textstyle{I=J\atop i_z\,s_z}\big\rangle}
\ &=\
{\textstyle\big\langle{I'=J'\atop i'_z\,s'_z}}
\big|D^{(1)}_{al}(\hat A)
\,\int_{SU(2)}dA\,\big|A\big\rangle
{\textstyle\big\langle{A}\big|{I=J\atop i_z\,s_z}\big\rangle}
\cr&=\
\int_{SU(2)}dA\,D^{(1)}_{al}(A)\,
(2J'+1)^{1/2}(-)^{J'-s'_z}D^{(J')*}_{-s_z',i'_z}(A^\dagger)
\cr&\qquad\qquad\times\
(2J+1)^{1/2}(-)^{J-s_z}D^{(J)}_{-s_z,i_z}(A^\dagger)
\cr&=\ \big[(2J+1)(2J'+1)\big]^{1/2}\,(-)^{J'-J+s_z+i_z'}\cr&\qquad\times\
\pmatrix{1&J&J'\cr a&i_z&-i'_z\cr}
\pmatrix{1&J&J'\cr l&-s_z&s_z'\cr}\ .}}
To obtain the first equality we have inserted a complete set of
$\ket{A}$ states on which $D^{(1)}_{al}(\hat A)$ is sharp; the
second equality follows from Eq.~\bracketdef; and the third, from
the textbook expression for the integral of three Wigner $D$-functions.
The two resulting $3j$ symbols express conservation of isospin
and angular momentum, respectively, while the overall square-root coefficient
embodies the proportionality rule \bf (v)\rm.

It is easily checked that the terms in Eq.~\PVdef\ involving the
spatial derivatives of the pion are correctly
reproduced by Eqs.~\PVAdef\ and \Ddef,
once one specializes to `in' and `out' nucleons by plugging in $J=J'=\hf$.
In contrast, the time derivative of the pion has been dropped in
moving from \PVdef\ to \PVAdef. This is because $\partial_0\vpi$ multiplies
the Dirac matrix $\gamma^0
\gamma^5$ which couples the `large' components of the baryon's Dirac
spinor to the `small' components, the latter being down by
$v/c\sim 1/N_c$. As often happens, the $1/N_c$
expansion has broken apart a Lorentz-invariant quantity.

Beyond the nucleons, the coupling \PVAdef\ contains useful phenomenological
information
about the higher $I=J$ baryons as well. Sandwiching it between nucleon and
$\Delta$ states using Eq.~\Ddef, one calculates
a $\Delta\rightarrow N\pi$ decay width within a few
MeV of its experimental value of 120 MeV \refs{\ANW,\DHM,\DiakPet}.
 In the same way,
one discovers that the higher baryons have
widths so large ($\Gamma_{{5\over2}\rightarrow\Delta\pi}\,\approx\,
800\,$MeV,
$\Gamma_{{7\over2}\rightarrow{5\over2}+\pi}\,\approx\,2600\,$MeV,
$\Gamma_{{9\over2}\rightarrow{7\over2}+\pi}\,\approx\,6400\,$MeV, etc.)
that they cannot sensibly be regarded as ``particles'' at
all \refs{\DHM,\DiakPet}. So
these higher-spin states, the existence of which is
often considered a major failing of the
large-$N_c$ approach, are actually \it not \rm in conflict with
phenomenology.\foot{The story of these unwanted large-$N_c$ baryons is
much more complicated in models with three or more light flavors. The
3-flavor Skyrme model \Guadagnini,
for example,  predicts exotic baryons not just with high spin but also
with \it low \rm spin, for instance a spin-$\hf$ antidecuplet.
In quark-model language such exotics map onto
baryons with $N_c$ quarks plus extra $q\bar q$ pairs \Manoharold.
Large-$N_c$ counting implies that each such pair costs a factor of
$1/\sqrt{N_c}$ to produce \refs{\tHooft,\Witten}. And indeed
one can show that matrix elements of physically relevant
operators between ``normal'' and ``exotic'' baryons are down precisely by
$1/N_c^{k/2}$ where $k$ is the number of such $q\bar q$ pairs; see
Ref.~\DJMnew, Sec.~XIII for details.}

Like the pions,
the allowed nucleon couplings of the $\rho,$ $\omega$ and/or $\sigma$ mesons
(for example) must also be echoed by couplings to the entire $I=J$
tower \action:
\eqna\mesoncouplings
$$\eqalignno{
&g_{\rho\NN}^\bare\partial_\mu\vrho_\nu\cdot\bar N\sigma^{\mu\nu}\vtau N
\ \longrightarrow\
3g_{\rho\NN}^\bare\epsilon_{ijk}
{\partial\over\partial {x^i}'}\rho^a_j(x)\,D^{(1)}_{ak}\big(A(t')\big)
\,\delta^3(\bx')&\mesoncouplings a
\cr&
g_{\omega\NN}^\bare\omega_\mu\bar N\gamma^\mu N
\ \longrightarrow\
g_{\omega\NN}^\bare\omega^0(x)\,\delta^3(\bx')&\mesoncouplings b
\cr&
g_{\sigma\NN}^\bare\sigma\bar NN
\ \longrightarrow\
g_{\sigma\NN}^\bare\,\sigma(x)\,\delta^3(\bx')
\quad,&\mesoncouplings c}$$
again dropping Dirac structures that involve `small' components. The
$N_c$ scalings are
\eqn\Ncscalings{\gpnnbare\sim g^\bare_{\rho\NN}
\sim g^\bare_{\omega\NN}\sim g^\bare_{\sigma\NN}\sim\sqn\ .}
Conspicuously absent from this
 list of permitted Yukawa couplings are the \it vector \rm
coupling of the $\rho$ meson and the \it tensor \rm coupling of the $\omega$
meson, $g^{\rm vec}_{\rho\NN}\vrho_\mu\cdot\bar N\gamma^\mu\vtau N$ and
$g^{\rm tens}_{\omega\NN}\partial_\mu\omega_\nu\bar N\sigma^{\mu\nu}N$.
These alternative couplings are forbidden by rule \bf (vi) \rm given
earlier, meaning that in contrast to \Ncscalings,
\eqn\Ncscalingsbad{g^{\rm vec}_{\rho\NN}\sim
g^{\rm tens}_{\omega\NN}\sim{1\over\sqn}}
and we can forget about them. The incorporation of these additional mesons
is deferred to future work;\foot{A natural extension of the ideas in
the present paper, and one more in keeping with Wilson's original
philosophy of the renormalization group \KogWil,
 is to add more and more mesons
to the model as the cutoff $\lam$ is increased. This is consistent with
the spectral representation of large-$N_c$ physics, which
requires, not just an infinite tower of baryons, but apparently
also an infinite number of mesons in each channel  \Witten. The smearing
of the baryon over a distance $\laminv$ can then be thought of as being due
to the interactions with mesons of mass greater than $\lam,$ which are
omitted from the model, whereas all mesons of mass less than $\lam$ are
explicitly kept.}
 for simplicity the explicit meson models
analyzed below will be built from pions alone.

\subsec{Constructing $\Lseagull$}

Finally we discuss couplings such as Fig.~3a, in which more than one
meson interacts with the baryon at the same space-time point. For want
of a better word we refer to these vertices  as ``seagulls.''
{}From QCD one can show that $n$-meson seagulls generically
scale like $N_c^{1-n/2}$  (rule \bf (ii) \rm above).
A familiar example from chiral perturbation theory is the coupling to
the nucleon of the pion's axial current, which is formed from the $U$
field, $U=\exp\big(2i\vpi\cdot\vtau/\fpi\big).$ Taylor expanding $U$
gives coefficients proportional to $g_A^{}\fpi^{-n}$ for the $n$-pion
couplings to the nucleon, where $n=1,3,5,\cdots$. Since $g_A^{}\sim N_c$ and
$\fpi\sim\sqrt{N_c}$ these indeed have the full-strength scaling behavior
with $N_c.$ An exception to this scaling rule is the set of
$n$-pion seagulls arising from
the \it vector \rm current coupling. These are proportional to
$g_V^{}\fpi^{-n}$, $n=2,4,6,\cdots,$
and since $g_V^{}=1$ by vector current conservation
they all drop out as $N_c\rightarrow\infty.$

Quite aside from any ``bare'' seagulls  that one
may choose to include  in the bare Lagrangian, one must also examine the
\it effective \rm seagulls that arise from the backwards-in-time
baryons ($Z$-graphs), once a Feynman diagram is decomposed into a sum of
time-ordered diagrams.
 Inevitably, these approximately pointlike induced vertices are termed
``$Z$-gulls.'' As illustrated in
Figs.~3b-c, they come from approximating the $V$-spinor
propagator by the constant $-i/2\Mren$ up to nonlocal ${\cal O}(N_c^{-2})$
corrections, where $\Mren$ is the physical baryon mass.
 Naively, the strength of the $Z$-gull
in Fig.~3c is then  $\gbare^2/\Mren\sim N_c^0$; since this is the same order
as a full-strength 2-meson bare seagull, it appears such a vertex
must be kept.
However, this naive counting is \it incorrect\rm.\foot{We are indebted to
Jim Friar for pointing out an error in an earlier draft of this paper.}
The reason is easily seen by writing out the Feynman rules for the $U$
and $V$ spinors directly.
For precisely those Yukawa couplings
\PVAdef\ and \mesoncouplings{} permitted by the $I_t=J_t$ rule \bf(vi)\rm,
there is an extra suppression of $\vec\sigma\cdot{\bf p}/M\,\sim\,
1/N_c$ at each of the two vertices in Fig.~3b. This is the cost
of turning a $U$-spinor into a $V$-spinor or \it vice versa\rm,
when the meson coupling is dominantly block-diagonal in the Dirac space.
We conclude that in self-consistent large-$N_c$ models, $Z$-gulls are actually
suppressed by two powers of $1/N_c$ compared with bare seagulls, and
may safely be dropped.

For simplicity, in the explicit models analyzed below,
 we will choose to  set to zero all bare seagulls as well.
Instead we will focus on the renormalization of the Yukawa
interaction and show that, in many cases, it corresponds to an irrelevant
operator of the large-$N_c$ renormalization group. We conjecture
that in these cases
the higher-point bare seagull couplings are  also irrelevant
operators, but we will leave this interesting question for future work.

\newsec{Large-$N_c$ Semiclassical Analysis}
\subsec{Formal summation of the leading-order graphs}
\def\mpi{m_\pi}
\def\picl{\pi_\cl}
\def\vpicl{\vpi_\cl}
\def\stat{{\rm stat}}
\def\pistat{\pi_\stat}
\def\vpistat{\vpi_\stat}
We now return to the central problem posed
in Sec.~1, namely the summation of the
leading-order contributions to the renormalized
Yukawa constants $g_\ren.$ Recall from Fig.~2 that the leading-order
graphs are those containing no purely mesonic loops, in other words,
those graphs which would be meson trees if one were to erase the
baryon line. The complete set of such graphs is
captured in Fig.~4.

Not surprisingly, being tree-like, Fig.~4 can be generated as the
solution to a classical equation of
motion. As a concrete example, with pions only, suppose that
\eqn\Lpidef{\Lmeson\ \equiv\
L_\pi\ =\ \hf(\partial_\mu\vpi)^2-\hf\mpi^2\vpi^2-V(\vpi)}
where $V$ contains the quartic and higher pion self-interactions. The
pion trees with one external pion line sum to a quantity we call
$\vpicl$ (the subscript `cl' standing interchangeably for `classical'
or `cloud') which solves the Euler-Lagrange equation implied by
Eqs.~\Lpidef\ and \PVAdef:
\def\bx{{\bf x}}
\eqn\euler{(\square+\mpi^2)\picl^a(\bx,t)+{\partial V\over\partial\picl^a}
\ =\ 3\gpnnbare D^{(1)}_{al}\big(A(t')\big){\partial\over
\partial {x^l}'}\delta^3(\bx')\ ,}
$x'$ referring to the center-of-mass frame of the moving baryon as
before.
This equation is illustrated in diagrammatic form in Fig.~5. Comparing
Fig.~5 with Fig.~4, we find apparent agreement, save for the ``missing''
sum over the $n!$ tanglings. But this sum is \it already implicit \rm in
Fig.~5. To see this, insert into the right-hand side of
Fig.~5 the resolution of unity into
a sum over all $n!$ (boosted) time orderings of the attachment points $z_k$,
\eqn\unity{1\ =\ \sum_{\rho\in S_n}
\theta\big(z^{\prime\,0}_{\rho(2)}-z^{\prime\,0}_{\rho(1)}\big)
\theta\big(z^{\prime\,0}_{\rho(3)}-z^{\prime\,0}_{\rho(2)}\big)
\times\cdots\times
\theta\big(z^{\prime\,0}_{\rho(n)}-z^{\prime\,0}_{\rho(n-1)}\big)\ ,}
$\rho$ being a permutation,
and for each element in this sum relabel $z_{\rho(k)}\rightarrow z_k.$
In this way we recapture, not Fig.~4 precisely, but rather Fig.~6,
which differs from Fig.~4 only in the time-ordering prescription up the
baryon line. This difference is truly unimportant; backwards-in-time
baryon propagation is always a $1/N_c$ effect.\foot{In this particular
instance the suppression of $Z$-graphs is even \it greater \rm than
that discussed in Sec.~2.5 above; shrinking a $V$-spinor propagator
to a point here results in a meson loop, which itself is $1/N_c$
suppressed (see Fig.~2d).}

To summarize, Eq.~\euler\ correctly accounts for the leading-order
contributions to the renormalized meson-baryon coupling, up
to $1/N_c$ corrections. The answer is conveniently expressed as a
\it quantum mechanical \rm path integral over the baryon's translational
and (iso)rotational collective coordinates:
\eqn\FPI{\int\calD A(t)\calD\bX(t)\,\picl^a(\bx,t)\,\exp i\int dt
\Big(\Lbaryon+\Lmeson[\vpicl]+\Lyukawa[\vpicl]\,\Big)\ ,}
ignoring seagulls for simplicity as stated above.
Our progress to this point has been that the path integration over the
field-theoretic variable $\vpi$ has been carried out in semiclassical
approximation. This simply means replacing $\vpi$ by $\vpicl$ everywhere
in \FPI, the justification being that $1/N_c$ always appears in the
combination $\hbar/N_c$ as highlighted earlier.

Thanks to the rescaling argument of Sec.~2.3, the meson cloud, like a
skyrmion, is  a nonperturbatively large configuration, scaling
like $\fpi\sim\sqrt{N_c},$ although its spatial extent goes
like $N_c^0.$ Another general
 point: it is not sufficient that the cloud be a solution
to the Euler-Lagrange equation; it must actually be a local \it minimum\rm,
in other words it must be  locally stable  against
small deformations. This stability issue will reemerge in Sec.~8
below. The interesting related question, What does an unstable cloud
collapse \it into\rm?, is examined at the end of the Appendix.

\subsec{Baryon self-energy and baryon-meson vertex corrections}
What is the meaning of the terms $\Lmeson[\vpicl]+\Lyukawa[\vpicl]$
in Eq.~\FPI?
Graphically, the answer can be seen in Fig.~7. As Fig.~7c in particular
makes clear, these terms fully account for the meson-tree baryon
self-energy and baryon-meson vertex corrections that we have neglected
till now \refs{\Luty,\Jenkins}.
 Taylor-expanding the exponential of these terms produces an
arbitrary number of such insertions at all placements along the
baryon line, automatically with the correct combinatorics.
Moreover, since $\vpicl$ itself depends on the baryon collective
coordinates $\bX(t)$ and $A(t)$ through the Yukawa source  on the right-hand
side of \euler, $\Lmeson[\vpicl]+\Lyukawa[\vpicl]$ have this dependence
as well, and on general grounds must have the form\foot{The third
term on the right-hand side, taken together with the identity
\schulmanid, establishes for \it all \rm leading-order baryon self-energy and
baryon-meson vertex corrections, and for \it any \rm  value of the pion mass,
 the  self-consistency of the hyperfine baryon mass splittings
 originally noted by Jenkins for the simplest such
graph in the chiral limit \Jenkins. As for the first two terms
on the right-hand side,  the fact that the same $\Mcl$ appears twice,
 with a relative weighting of $-\hf\dot\bX^2,$
 follows trivially from Lorentz invariance (as Jim Hughes has
reminded us). So long as
the Lagrangian \it density \rm is a Lorentz scalar, meaning that
${\cal L}(x)\rightarrow{\cal L}(x')$ as $x\rightarrow x',$ then
$L(t) \equiv \int d^3\bx\,{\cal L}(x)
\rightarrow \int d^3\bx\,{\cal L}(x')
=\sqrt{1-\dot\bX^2}\,L(t')
= \big(1-\hf\dot\bX^2+{\cal O}(\dot\bX^4)\big)L(t'),$ where
$\sqrt{1-\dot\bX^2}$ is the change-of-frames Jacobian.
One \it cannot \rm ignore the Lorentz contraction of the cloud, as
the two masses would then be unequal.}
\eqn\cloudaction{\Lmeson[\vpicl]+\Lyukawa[\vpicl]\ =\
-\Mcl+\hf \Mcl\dot\bX^2+\Icl\Tr\dot A^\dagger\dot A+\cdots\ .}
Since $\vpicl\sim\sqrt{N_c}$ it follows that
$\Mcl$ and $\Icl$ scale like $N_c$, just like $\Mbare$ and $\Ibare.$
The dots in Eq.~\cloudaction\
indicate terms suppressed in the $1/N_c$ expansion,
such as $\dot\bX\dot A$ cross terms.\foot{To prove that such
terms are suppressed one needs to anticipate
 the findings of Sec.~3.3 below, and check that when the classical
meson fields are precisely hedgehogs the cross terms vanish
by symmetry upon spatial integration. They are only nonvanishing to the
extent that the cloud deviates from the hedgehog ansatz, which it
does at a higher order in $1/N_c$ \refs{\DHM,\Schroers}.}

The lesson of Eq.~\cloudaction\
 is that the classical meson cloud $\vpicl$ (likewise
$\vrho_\cl,$ $\omega_\cl,$ $\sigma_\cl,$ etc.) gives a \it form-preserving
\rm and {\it leading-order}\foot{This contrasts with the quantum corrections
to the mass of the Skyrmion, which are only ${\cal O}(N_c^0).$ This is
because the Skyrmion, unlike the ``bare'' nucleon, is the solution to
an Euler-Lagrange equation.} correction to the `bare' baryon expressions
\lbaryondef-\MbareJdef, so that effectively
\eqn\MrenJdef{\Mbare^J\ \longrightarrow\ \Mren^J=\ \Mren+{J(J+1)\over2\Iren}}
where the renormalized baryon mass and moment of inertia are simply
the sums\foot{The reader might be confused about the present
definition of mass renormalization, versus the use of
 a conventional mass counterterm $\delta M=\Mbare-\Mren$.
 In fact, they are the same thing. By definition,
$\delta M$ must be tuned to cancel the shift in the pole of the propagator
away from $\Mren$ induced by the interactions with the mesons.
Since the nonrelativistic propagator $i/( k_0-M+i\epsilon)$
Fourier transforms to $\theta(\Delta t)e^{-iM\Delta t},$ this simply
means that $-i\delta M\Delta t$ must cancel the meson cloud contribution
$-i\Mcl\Delta t$ to the effective action.
But this condition $\delta M=-\Mcl$ is just a rewrite of
%danger
Eq.~(3.7), \sl QED\rm.}
\eqn\Mrendef{\Mren=\Mbare+\Mcl\ ,\qquad\Iren=\Ibare+\Icl\ .}
\subsec{Hedgehog meson clouds}
We now examine more carefully the structure of the classical meson cloud.
Once again a helpful example is the
``pions-only'' Lagrangian \Lpidef. A simpler recasting of the Euler-Lagrange
equation \euler\ comes from nailing the baryon's center of mass at
the origin and its (iso)spin orientation at the North pole, $\bX(t)=0$
and $A(t)=1,$ in which case \euler\ becomes \ArnMat
\eqn\stateuler{(-\nabla^2+\mpi^2)\pistat^a(\bx)+{\partial V
\over\partial\pistat^a}\ =\ 3\gpnnbare{\partial\over
\partial x^a}\delta^3(\bx)\ .}
The solution is termed the ``static pion cloud'' $\vpistat(\bx),$ whereupon
$\vpicl$ is approximated by
\eqn\piclpistat{\picl^a(\bx,t)\ \cong\ D^{(1)}_{al}\big(A(t')\big)
\pistat^l(\bx')}
up to $1/N_c$ corrections,
 the primes denoting the baryon's
center-of-mass frame as always.
 So it will suffice to focus on Eq.~\stateuler\
rather than the more complicated time-dependent equation
\euler. In terms of $\vpistat,$ the quantities
$\Mcl$ and $\Icl$ are explicitly given by
\eqn\MclIcl{\Mcl\ =\ \int d^3\bx\,\Big[
\hf\big(\partial_i\vpistat\big)^2+\hf\mpi^2\vpistat^2+V(\vpistat)
-3\gpnnbare\vpistat\cdot\nabla\delta^3(\bx)\,\Big]\ }
and
\eqn\IclMcl{\Icl\ =\ {\textstyle{2\over3}}\int d^3\bx\,\vpistat^2\ .}

What does $\vpistat$ look like? An important hint is that the index
`$a$' lives in \it iso\rm space on the left-hand side and ordinary space
on the right-hand side of Eq.~\stateuler. Therefore,
as in the Skyrme model, the solution can generically
be found in the maximally symmetric ``hedgehog ansatz,''
 which equates these two spaces \ArnMat:
\eqn\hedgehog{\vpistat(\bx)\ =\ {\fpi\over2}\hat\bx\,F(r)}
where $r=|\bx|$ and $\hat\bx=\bx/r.$ In turn, the cloud profile $F(r)$
solves a model-dependent nonlinear ordinary differential equation in the radial
coordinate $r$, obtained by plugging \hedgehog\ into \stateuler.

Even if additional mesons are incorporated as per Eq.~\mesoncouplings{},
 the coupled static equations
(one for each meson species) are still solved in the hedgehog ansatz,
suitably generalized in the manner familiar from vector-meson-augmented
Skyrme models \refs{\ninej,\vecmesrefs}:
\eqn\vecmes{\eqalign{&\rho^{ai}_\stat(\bx)=\fpi\epsilon_{iak}\hat x^k
G(r)\ ,\quad\rho^{a0}_\stat(\bx)=0\ ,\quad\sigma_\stat(\bx)=
\fpi H(r)\ ,\cr&
\omega^i_\stat(\bx)=0\ ,\quad
\omega^0_\stat(\bx)=\fpi K(r) ,\quad\hbox{etc.,}}}
where `$i$' and `$a$' label spin and isospin, respectively.
Since $\fpi\sim\sqn$ has been factored out explicitly, the profiles
$F$, $G$, $H$ and $K$ each scale like $N_c^0.$ In general they
obey model-dependent coupled nonlinear radial ODE's.

\def\gpnnren{\gpnn^\ren}
\subsec{Ultraviolet divergences and the large-$N_c$ renormalization group}
The above discussion contains something of a cheat: as written,
 Eq.~\stateuler\ only admits a mathematically well-defined solution in
the free-pion case $V(\vpi)\equiv0,$ in which case obviously
\eqn\freepions{\vpistat(\bx)\ \propto\ \vec\nabla\,{e^{-\mpi r}\over r}\ .}
For nonvanishing $V$, the lack of a solution to this nonlinear equation
is due to the exact pointlike
nature of the $\delta$-function source.\foot{This arcane technical point
about the lack of a well-defined solution
is confirmed in the explicit examples of Secs.~5-8 below, as the
reader can verify by imagining the following exercise. Hold $\gpnnbare$
 fixed  (instead of $\gpnnren$ as in the large-$N_c$ renormalization
group) while taking the ultraviolet cutoff $\lam\rightarrow\infty$,
and confirm that in so doing
$\vpistat$ has either a singular limit, or no limit at all.}
 So, too, the cloud parameters
$\Mcl$ and $\Icl$ introduced in Eq.~\cloudaction\ are actually
ill-defined, diverging in the ultraviolet. An ultraviolet cutoff
is required. This should come as no surprise: in most hadron models the
need for such a cutoff arises as early in the discussion as the first loop
correction in the original Feynman graphs (e.g., Figs.~2b-c). But
even if one were to concoct a meson model free of this type of
divergence at the level of the original graphs, a cutoff
$\lam\sim N_c^0$ on the meson momenta
would \it still \rm be required for the self-consistency
of the subsequent formalism; specifically, one needs to ensure that
the baryon always stays in the vicinity of its rest frame even after
an arbitrary number of interactions with the mesons. In fact,
if instead one were to permit meson momenta of order $N_c,$ in
particular above the $N\bar N$ threshold, it is plausible that the
entire effective Lagrangian approach breaks down (see Ref.~\Witten,
Sec.~8.3).

The simplest fix, with interesting consequences as we shall see,
is to smear out the $\delta$-function in some manner,
\def\deltalam{\delta_\lam}
\eqn\smear{\delta^3(\bx)\ \longrightarrow\ \deltalam^3(\bx)\ ,}
over a length scale $\laminv.$ While details of this regulator should
not matter for sufficiently large $\lam,$ we will nevertheless
insist that $\deltalam^3
(\bx)$ be spherically symmetric, so that the hedgehog ansatz remains valid
even for finite $\lam$, an enormous technical simplification.

Now suitably regulated, Eq.~\stateuler\  (returning
to the pions-only example) is easily solved numerically for $F(r)$.
The solution is shown schematically
in Fig.~8. For $r$ well inside the classical cloud, whose radius
is determined by the parameters of $V$, the behavior of $F(r)$ is highly
model-dependent. But
 for $r\gg\mpi^{-1}$, Eq.~\stateuler\ linearizes, and one finds
\eqn\Fasym{F(r)\ \longrightarrow
\ {3\gpnnren\over2\pi\fpi }\,\Big({1\over r^2}+{\mpi\over r}\Big)
e^{-\mpi r}}
which follows from \hedgehog\ and \freepions. Notice the
 new parameter, $\gpnnren,$ which measures the height of the
exponential tail. While its precise numerical value depends
sensitively on the choices of $V(\vpi)$ and $\lam,$ its physical
interpretation as the renormalized pion-nucleon pseudovector
coupling constant is
pleasingly model-independent. To our knowledge this identification was first
made in Sec.~4 of Adkins, Nappi and Witten \ANW,
 and has recently been confirmed
in Ref.~\DHM\ with a careful analysis of the LSZ amputation
procedure.\foot{While both these references discuss this particular
issue in the context of the Skyrme model, the reader can verify
that the conclusions are equally valid for effective Lagrangian models
such as concern us here, with explicit baryon sources. It should come
as no surprise that the LSZ residue is sensitive only to the asymptotic
behavior \Fasym, as this is a well-known property of Fourier transforms.
The fact that the renormalized coupling is still pseudovector, like
the bare coupling, is
simply because the Fourier transform of a hedgehog is necessarily
proportional to $\bf q$.
The only subtle feature is this \DHM:
 the existence of an LSZ pole precisely on the pion
mass shell depends crucially on a small quadrupole \Schroers\
deviation of the Skyrmion (or, in the present context, meson cloud)
 \it away \rm from the hedgehog ansatz, induced by the
baryon's (iso)rotations. This distortion is one of the $1/N_c$ corrections
dropped in Eq.~\piclpistat, as it is tangential to our present
purposes.} The behavior \Fasym\ is equally valid in the chiral limit:
\eqn\Fasymchiral{F(r)\ \longrightarrow
\ {3\gpnnren\over2\pi\fpi r^2} \qquad\hbox{(massless pions)}\ .}

In summary, for any given choice of meson Lagrangian, we have described an
explicit numerical procedure---Eqs.~\Mrendef\ and \Fasym---for
 extracting the physical, renormalized
parameters $\Mren,$ $\Iren$ and $\gpnnren$ as functions of the Lagrangian input
quantities $\Mbare,$ $\Ibare$, $\gpnnbare$ as well as the UV cutoff $\lam.$
Alternatively, one might wish to \it fix \rm
$\Mren,$ $\Iren$ and $\gpnnren$, say
to their experimental values
\eqna\pdb
$$\eqalignno{&\Mren+{3\over8\Iren}\equiv M_N\cong939\,\hbox{MeV}\ ,
&\pdb a
\cr
&\Mren+{15\over8\Iren}\equiv M_\Delta\cong1232\,\hbox{MeV}\ ,
&\pdb b
\cr
&\gpnnren\,\cong\,{13.5\over2M_N}\ ,
&\pdb c}$$
then solve implicitly for $\Mbare(\lam),$ $\Ibare(\lam)$ and
$\gpnnbare(\lam).$ This latter approach seems the most reasonable
to us, and will be our philosophy from
now on. We call this novel program the \it large-$N_c$
renormalization group\rm, and devote the remainder of this paper to
exploring its solutions  in a variety of illustrative models.

\newsec{Free pion Lagrangian and its continuum limit}

Our first example consists simply of free massless pions,
\eqn\Lfree{\calLmeson\ =\ \hf(\partial_\mu\vpi)^2\ ,}
coupled derivatively to the $I=J$ baryon tower as per Eq.~\PVAdef. In the
hedgehog ansatz \hedgehog, the static Euler-Lagrange equation \stateuler\
becomes
\eqn\Feqn{F''+{2\over r}F'-{2\over r^2}F\ =\ 6\fpinv
\gpnnbare(\lam){\partial\over\partial r}
\deltalam(r)\ .}
This being a linear equation, it
 is trivially solved using the method of Green's functions:
\eqn\Fsolve{F(r)\ =\  6\fpinv\gpnnbare(\lam)
\int_0^\infty dr'\,{r'}^2\,G(r,r')\,{\partial\over\partial r'}
\deltalam(r')\ ,}
where the massless Green's function that is well behaved at both $r=0$ and
$r=\infty$ is
\def\rless{r_<}
\def\rgreater{r_>}
\eqn\Gdef{G(r,r')\,=\,-{\rless\over3\rgreater^2}\ ,\quad
\rless=\min\{r,r'\},\quad\rgreater=\max\{r,r'\}\ .}

The renormalized
 Yukawa coupling $\gpnnren$ is extracted from the large-distance
behavior of $F$ as per Eq.~\Fasymchiral. With the mild (and relaxable)
assumption that $\deltalam(r')$ has compact support, Eqs.~\Fsolve\
and \Gdef\ imply
\eqn\freeFasym{F(r)\ \matrix{&{\scriptstyle r\rightarrow\infty}
\cr&\longrightarrow\cr&{}}
\quad {6\gpnnbare(\lam)\over\fpi r^2}
\int_0^\infty dr'\,{r'}^2\,\deltalam(r')\ =\
{3\gpnnbare(\lam)\over2\pi\fpi r^2}\ .}
We have made use of an integration by part, plus the requirement that
the volume of $\deltalam$ be normalized to unity (regardless of other
details of this smearing function).
Comparing Eqs.~\freeFasym\ and \Fasymchiral, we deduce
\eqn\equivalence{\gpnnbare(\lam)\ =\ \gpnnren }
for all $\lam,$
admittedly not a surprising result for free field theory, but a reassuring
sanity check on our formalism.

This lack of any flow is consistent with what might be termed
an ``exact'' Cheshire cat picture \refs{\Nielsen,\Rho}
 (recalling the discussion in Sec.~1).
In truth, this is the \it only \rm model we have found where this
perfect equality holds. For example,
the mildest conceivable modification to the Lagrangian \Lfree\
is to add a pion mass term. In that
case the Green's function is, instead,
\eqn\Gmassdef{G(r,r')\ =\ {1\over2\mpi^3}\Big[
\Big({1\over\rless^2}-{\mpi\over\rless}\Big)e^{\mpi \rless}-
\Big({1\over\rless^2}+{\mpi\over\rless}\Big)e^{-\mpi \rless}\Big]
\Big({1\over\rgreater^2}+{\mpi\over\rgreater}\Big)e^{-\mpi \rgreater}}
which properly reduces to \Gdef\ as $\mpi\rightarrow0,$ and \equivalence\ is
amended slightly to
\eqn\inequivalence{\gpnnbare(\lam)\ =\ \Big(1+{\cal O}(\mpi^2/\lam^2)
\Big)\gpnnren\ . }

 In either variation, massless or massive,
the continuum limit $\lam\rightarrow\infty$ can be safely taken,
and the ``ultraviolet fixed point'' that emerges is just what one started
with: a theory of free pions derivatively coupled to the baryon tower. This is
entirely expected, since as mentioned earlier $V(\vpi)\equiv0$ is the only
case in which Eq.~\stateuler\ as written already has a \it bona fide \rm
solution, namely, Eq.~\freepions, and there is no actual need to smear out the
$\delta$-function source. We now turn to more interesting
 examples where these statements no longer hold, and where
the breakdown of the ``exact'' Cheshire
cat picture is much more severe than Eq.~\inequivalence.

\newsec{The nonlinear $\sigma$ model and its {\it lack} of continuum limit}

For our second example, consider the nonlinear $\sigma$ model for
pions,
\eqn\NLSMdef{\calLmeson\ =\ {\fpi^2\over16}\,\Tr\,\partial_\mu U^\dagger
\partial^\mu U\ ,\quad U=\exp\big(2i\vpi\cdot\vtau/\fpi\big)\ ,}
again augmented by the bare Yukawa coupling \PVAdef. The static
Euler-Lagrange equation \stateuler\ now works out to
\eqn\FNLSM{F''+{2\over r}F'-{1\over r^2}\sin 2F\ =\
6\fpinv\gpnnbare(\lam){\partial\over\partial r}
\deltalam(r)\ .}
Solving this nonlinear equation for $F(r)$
 requires that we specify a smearing of the source.
For convenience, we follow Ref.~\ManPLB, and choose a radial step-function
\def\FI{F_{\sst\rm I}}
\def\FII{F_{\sst\rm II}}
\def\lineone{ {3\lam^3\over4\pi}\ :\quad r\le\laminv}
\def\linetwo{ 0\ \ :\quad r>\laminv}
\eqn\mansmear{\deltalam(r)\ =\
\Bigg\{\matrix{&\lineone\cr&\linetwo\cr}}
which is properly normalized to unit volume. The technical advantage,
which we exploit presently, is that
the right-hand side of Eq.~\FNLSM\ is now proportional to a \it
true \rm $\delta$-function, since
\eqn\manderiv{{\partial\over\partial r}\deltalam(r)\ =\
-{3\lam^3\over4\pi}\delta(r-\laminv)\ .}
There is also a conceptual advantage: the right-hand side of \manderiv\
means that the baryon and meson degrees of freedom only interact
at the ``bag radius'' $\laminv,$ which sharpens the analogy to the
traditional chiral bag \refs{\Chodos{--}\Rho}.

With this convenient choice of regulator,
 the prescription for satisfying Eq.~\FNLSM\ is
transparent: First solve the \it homogeneous \rm version of
Eq.~\FNLSM, namely
\eqn\homoNLSM{F''+{2\over r}F'-{1\over r^2}\sin 2F\ =\ 0\ ,}
for $r<\laminv$ (``region I'') and for $r>\laminv$ (``region II'')
subject to the boundary condition \Fasymchiral; next,
match the solutions in these two regions,
$\FI(r)$ and $\FII(r)$, at the point $r=\laminv$;
and finally, read off $\gpnnbare(\lam)$ from the slope
discontinuity,
\eqn\discont{\gpnnbare(\lam)\ =\ {\textstyle{2\over9}}\pi\fpi
\lam^{-3}\big(\FI'(\laminv)-\FII'(\laminv)\big)\ .}

This 3-step graphical procedure is illustrated in Fig.~9. The curves
$\FII(r)$ and $\FI(r)$ are displayed in Figs.~9a and 9b, respectively.
The curve $\FII(r)$ is uniquely specified by $\gpnnren$ through
the asymptotic formula \Fasymchiral. In contrast,
 $\FI(r)$ actually stands for an entire one-parameter
 family  of curves related by dilatations, $\FI(r)\rightarrow
\FI(\lambda r),$ thanks to the scale invariance of Eq.~\homoNLSM.
For any specified value of the cutoff $\lam,$
the scale parameter $\lambda$ needs to be tuned so that
 $\FI(\laminv)=\FII(\laminv).$ Note that $\FI,$ unlike $\FII,$
attains a maximum value $\FI^{\rm max}\cong.58\pi$
before dropping back down towards $\pi/2.$

Let us discuss the qualitative behavior of the
large-$N_c$ renormalization group as the cutoff
increases from zero. In the infrared regime
$\lam\,\ll\,(\fpi/\gpnnren)^{1/2}$, the flow of $\gpnnbare(\lam)$
necessarily approaches the free massless pion case, Eq.~\equivalence.
This is simply because the patched-together profile $F(r)$ is small
everywhere, in which case Eq.~\FNLSM\ reduces to Eq.~\Feqn\ up to
${\cal O}(F^3)$ corrections. This is precisely the regime studied
recently by Manohar \ManPLB, who correctly reproduced
the ${\cal O}(\mpi^3)$ non-analytic correction to the
baryon mass familiar from one-loop chiral perturbation theory.

But the behavior of the renormalization group for higher $\lam$  quickly
diverges from the free-pions example.
Notice that as $\lam$ passes a first critical value $\lam_1$ (the
point where $\FII(\lam^{-1}_1)=\pi/2$) a second
disconnected solution to Eq.~\FNLSM\
emerges, in which $\FII$ intersects $\FI,$ not in the branch of the
curve labeled ``Region A'' but rather in ``Region B.'' The flow in this
branch is still dictated by Eq.~\discont, but since for any given $\lam$
the value of $\FI'(\laminv)$ differs between the two branches, the
solutions are distinct.  As $\lam$ increases further,
these two branches of $\gpnnbare(\lam)$ gradually approach one
another, until at a new critical value $\lam_2$, defined by
$\FII(\lam^{-1}_2)=\FI^{\rm max}$,  they coalesce. This latter
scale is indicated by a cross in Fig.~9. From
Fig.~9a one reads off $\lam_1^{-1}\cong .75(3\gpnnren/2\pi\fpi)^{1/2}$
and $\lam_2^{-1}\cong .68(3\gpnnren/2\pi\fpi)^{1/2}$, or in conventional
 units,
\eqn\lamtwodef{\lam_1\ \cong\ 310\,\hbox{MeV}\ ,\qquad
\lam_2\ \cong\ 340\,\hbox{MeV}}
using the physical values \pdb{}. We calculate $\gpnnbare(\lam_2)
\cong.43\gpnnren.$ Crucially, for $\lam>\lam_2$,
if one insists that $\FI(0)=0,$ then there
is no way to match $\FI$ with $\FII,$ hence no solution
to Eq.~\FNLSM.\foot{Technically speaking, the two independent
branches of $\gpnnbare(\lam)$, which have coalesced at $\lam_2$, leave the
real axis and bifurcate into complex conjugate pairs for $\lam>
\lam_2$.}

Thus we  have exhibited two ``phases'' of the model (for want of a better
term),  the first being defined for $\lam\in
[0,\lam_2]$, and the second only for $\lam\in[\lam_1,\lam_2].$
The critical ``bag radius'' $\lam_2^{-1}\cong.6\,$fm
  is the UV scale beyond which the
``chiral bag'' cannot be formed from the  2-derivative pion action
\NLSMdef\ alone; higher-derivative terms must be added.
 Evidently this breakdown owes nothing to soft-pion
arguments as one is accustomed to---but follows solely from large-$N_c$
reasoning.

To be honest, for $\lam>\lam_2,$  one \it can \rm patch together a
solution---if one allows
the cloud to have nonzero winding number ({\it cf.}
Eq.~(6.5$a$) below). For example, there are four ``phases'' of the
model with winding number unity, which are constructed as follows.
Leave $\FII(r)$ the same as above, but let
$\FI(r)\rightarrow\FI(r)+\pi,$ exploiting a
discrete symmetry of Eq.~\FNLSM. There are two such
solutions, again depending on whether the intersection takes place in
``Region A'' or in ``Region B.'' The two remaining solutions are
generated, instead, by the discrete symmetry
$\FI(r)\rightarrow-\FI(r)+\pi.$ Likewise there are four phases of
the model for any positive winding number $n$, generated by\foot{We
suspect that these
higher winding-number clouds may be, like the hedgehog Skyrmion in the
$B=2$ sector \refs{\Skyrme,\Jackson},
 unstable to small deformations in the cloud \it away \rm from the
hedgehog ansatz, but we have not looked for any such deformations,
neither in the present model nor in the ones to follow.}
$\FI(r)\rightarrow\pm\FI(r)+n\pi.$  Each such phase exists only for
a finite interval in the cutoff $\lam.$ In particular---unlike the
 free-pions example---here there is no phase within
which one can take the continuum limit $\lam\rightarrow\infty.$
In other words, {\it this model lacks an ultraviolet fixed point}.

Of course, there is no reason whatsoever that an effective field theory need
have a continuum limit. It would be perfectly reasonable to fix $\lam$
at a finite value less than $\lam_2$
where the ``chiral bag'' still makes sense,
and to calculate, for example, the static properties of this hybrid nucleon,
$\grave{a}$ $la$ Adkins, Nappi and Witten \ANW. Nevertheless, it is
instructive to pose the question: In a generic theory, if the continuum
limit can in fact be taken, what type of UV fixed point might one
expect? In the free-meson examples of Sec.~4,  $\gpnnbare(\lam)$ runs to a
finite, nonzero value as $\lam\rightarrow\infty$.
But this behavior must be the exception rather than the rule. For, a nonzero
limiting value of $\gpnnbare(\lam)$ suggests that a
solution to  Eq.~\stateuler\ exists even for a non-vanishing right-hand
side with an exact $\delta$-function source. Except for the free-meson case
$V\equiv0,$ we have yet to discover a differential
equation where this is possible. Instead, it is far more plausible that
$\gpnnbare(\lam)$ is the coupling constant of
an ``irrelevant operator,'' and
 \it vanishes \rm in the ultraviolet (as suggested by
 the factor of $\lam^{-3}$ on the right-hand side of Eq.~\discont).
The resulting continuum theory would then be completely
independent of the baryonic degrees of freedom, an interesting
example of ``universality'' \KogWil.
In this event, Eq.~\stateuler\ admits a solution if (and only if?)
the meson Lagrangian  supports a nontrivial configuration
in the absence of a baryonic source---meaning
a soliton or Skyrmion, either topological or energetic.

\def\Ustat{U_\stat}
Viewed in this light, it is no surprise that the nonlinear $\sigma$ model
coupled to the baryon tower has no UV limit, as we have just learned.
Plausibly, this is because Eq.~\NLSMdef\ does not by
itself support a Skyrmion. The  reason is Derrick's famous ``no go''
theorem (a variant of the virial theorem) \Derrick:
if one  posits a static Euler-Lagrange solution $\Ustat(\bx),$ then the energy
functional
\eqn\Estatdef{E[\Ustat]\ =\ {\fpi^2\over16}\int d^D\bx\,\Tr\,
\partial_i\Ustat^\dagger\partial_i\Ustat\ ,}
rather than being stationary, can actually
 be lowered arbitrarily by a homogeneous rescaling
\eqn\rescaling{\Ustat(\bx)\longrightarrow\Ustat(\lambda\bx)\ ,\quad
E[\Ustat]\longrightarrow{1\over\lambda^{D-2}}\,E[\Ustat]\ ;}
therefore, no such solution can exist.\foot{This point was apparently
missed by Gervais and Sakita \GerSak,
 whose Eqs.~(5.19)-(5.20) admit no solution.}
There are several known ways to modify the nonlinear $\sigma$
model to prevent such a
``Derrick collapse.'' One way, which we examine in the following
Section, is simply to reduce the dimensionality of space from $D=3$ to
the ``critical dimension''
$D=2,$ as Eq.~\rescaling\ suggests. As a bonus, the resulting
 toy model turns out to be analytically
soluble. Alternatively, and more physically, in Secs.~7 and 8 we will
augment the Lagrangian \NLSMdef\ by the 4-derivative ``Skyrme term,'' and
explore the interesting, and unexpected, consequences.

\newsec{An exactly soluble 2+1 dimensional model with a nontrivial
UV fixed point}

Motivated by the above discussion,
we would like to construct a 2+1 dimensional model which parallels as closely
as possible  the nonlinear $\sigma$ model of the previous
Section. To make the analogy as plain as possible,
 it is helpful to recall an alternate
parametrization of the pion field to that given in Eq.~\NLSMdef.
Rather than $U=\exp\big(2i\vpi\cdot\vtau/\fpi\big),$ take
\def\bu{{\bf u}}
\def\bn{{\bf n}}
\eqn\newUdef{U=u_0+i\bu\cdot\vtau\ ,\quad u_0+\bu^2=1\ ,\quad
\vpi={\fpi\over2}\,\bu\ ,}
in terms of which the nonlinear $\sigma$ model Lagrangian \NLSMdef\ is simply
\eqna\newNLSM
$$\eqalignno{&{\cal L}^{\rm 3D}_{\rm meson}
\ =\ {\fpi^2\over8}\big[(\partial_\mu u_0)^2
+(\partial_\mu u_1)^2+(\partial_\mu u_2)^2+(\partial_\mu u_3)^2
\big]\ .&\newNLSM a}$$
The natural 2+1 dimensional analog is then\foot{The mesonic sector
of this toy model is a simplified version of the ``baby Skyrme model'' of
Ref.~\baby.}
$$\eqalignno{&{\cal L}^{\rm 2D}_{\rm meson}
\ =\ {f^2\over8}\big[(\partial_\mu n_0)^2
+(\partial_\mu n_1)^2+(\partial_\mu n_2)^2
\big]\ ,\quad n_0^2+n_1^2+n_2^2=1\ .&\newNLSM b}$$
Just as the choice of vacuum $(u_0,\bu)=(1,{\bf0})$ spontaneously breaks
the $O(4)$ chiral symmetry of \newNLSM{a} down to isospin $O(3)$, so too
the vacuum choice $(n_0,\bn)=(1,0,0)$ in \newNLSM{b} breaks ``chiral''
$O(3)$ down to ``isospin'' $O(2)$.

The hedgehog ansatz for the pion cloud,
\eqna\hedgethreeD
$$\eqalignno{&(u_0,\bu)\ =\ \big(\cos F(r)\,,\,\hat\bx\sin F(r)\big)\ ,
&\hedgethreeD a}$$
will prove equally applicable to the lower-dimensional model,
$$\eqalignno{&(n_0,\bn)\ =\ \big(\cos F(r)\,,\,\hat\bx\sin F(r)\big)\ .
&\hedgethreeD b}$$
Evaluated on these ans$\ddot{\rm a}$tze, the Hamiltonians are quite similar:
\eqna\HamhedgethreeD
$$\eqalignno{&H_{\rm 3D}[F]\ =\
{\fpi^2\over8}\int d^3\bx\Big({F'}^2+2\,{\sin^2F\over r^2}\Big)
&\HamhedgethreeD a}$$
versus
$$\eqalignno{&H_{\rm 2D}[F]\ =\ {f^2\over8}
\int d^2\bx\Big({F'}^2+{\sin^2F\over r^2}\Big)\ ,
&\HamhedgethreeD b}$$
respectively. But there is a major difference in
the winding number formulae between the two models:\foot{Winding number
is given, respectively, by $W=-{1\over24\pi^2}
\int d^3{\bf x}\,\epsilon_{ijk}
\Tr\, T_iT_jT_k$ in three dimensions, where $T_i=U^\dagger\partial_iU$, and
$W={1\over8\pi}\int d^2\bx\,\epsilon_{\nu\mu}\,n\cdot\partial_\mu n
\times\partial_\nu n$ in two dimensions, where $n$ means $(n_0,n_1,n_2)$.}
\eqna\windingthreeD
$$\eqalignno{&W_{\rm 3D}[F]\ =\ -\int d^3\bx\,{F'\sin^2F\over2\pi^2r^2}
\ =\ {1\over\pi}\big[F(0)-F(\infty)\big]&\windingthreeD a}$$
versus
$$\eqalignno{&W_{\rm 2D}[F]\ =\ -\int d^2\bx\,{F'\sin F\over4\pi r}
\ =\ -\hf\big[\cos F(0)-\cos F(\infty)\big]\ .&\windingthreeD b}$$
In other words, whereas in three spatial dimensions
the hedgehog ansatz is broad enough to encompass
all integer winding numbers, the same is
\it not \rm true in two dimensions where winding number
is restricted to the three
values $\{-1,0,+1\}$ as follows from Eq.~\windingthreeD{b}.\foot{A mild
extension of the hedgehog ansatz \hedgethreeD{b} covers the sectors
with winding number $n$: rewrite $\hat\bx$ as $(\cos\theta,\sin\theta)$
and replace this by $(\cos n\theta,\sin n\theta)$ instead.}
This caveat is irrelevant for present purposes, and so we press on.

There is a well-known rewrite of the $O(3)$ model in terms of the
``conformal variables'' \Othree
\eqn\confvar{w\,=\,{n_1+in_2\over1+n_0}\ ,\quad
\bar w\,=\,{n_1-in_2\over1+n_0}\ .}
These variables (canonically rescaled by a factor of $f/2$)
are closer in spirit to the ``old'' pion representation
\NLSMdef, in that there are no extraneous fields such as $u_0$ or
$n_0$ that need to be eliminated with a spherical constraint. Paralleling
the previous Section, we will therefore use the $w$'s, not the $n$'s, in
constructing the Yukawa coupling.

Next we turn to the baryons. The obvious ``toy'' analog of the $I=J$ tower
is an infinite sequence of states $\ket{\nu}$ that transform as $(\nu,\nu),$
$\nu=\pm\hf,\pm\thf,\pm\fhf,\dots,$ under ``isospin'' $O(2)\cong U(1)$ and
spatial $U(1)$ rotations in the $x$-$y$ plane. A spin$\,\times\,$isospin
 invariant Yukawa coupling in the baryon rest frame then has the form
\eqn\toyuk{{f\over2}
\,{\partial w\over\partial z}\sum_{\nu=\pm\hf,\pm\thf,\cdots}
g_\nu\,\ket{\nu+1}\bra{\nu\,}
\ +\ \hbox{H.c.}}
Here the $g_\nu$ are arbitrary complex constants, $z=x+iy$,
and the operator $\ket{\nu+1}\bra{\nu\,}$ simply means that the
difference of the $U(1)$ (iso)spin charges between the initial and
final baryons must be unity.

An intelligent way to generate such a coupling is to work in the collective
coordinate basis $\ket{\theta},$ $\theta\in U(1),$ analogous to
$\ket{A}$  in 3+1 dimensions. The baryon Lagrangian analogous
to \lbaryondef\ is
\eqn\lbaryontoy{{\cal L}_{\rm baryon}^{\rm2D}
\ =\ -\Mbare+\hf\Mbare\dot\bX^2+\hf\Ibare\dot\theta^2
\ ,}
the last term representing free motion on the $U(1)$ manifold.
Another of Schulman's path integral identities \Schulman,
\eqn\schulmantoy{\eqalign{&\int_{\theta(t_1)=\theta_1}^{\theta(t_2)=\theta_2}
\calD \theta(t)\,\exp
i\int dt\big(-\Mbare+\hf\Ibare\dot\theta^2\big)\ =
\cr
&\sum_{\nu=\pm\hf,\pm\thf,\cdots}
\big\langle \theta_2\big|\nu\big\rangle
\,e^{-i(t_2-t_1)\Mbare^\nu}\,
\big\langle \nu\big|\theta_1\big\rangle  }}
with
\eqn\Mbarenudef{\Mbare^\nu\ =\ \Mbare+{\nu^2\over2\Ibare}\quad
\hbox{and}\quad
\big\langle \nu\big|\theta\big\rangle \ =\ e^{i\nu\theta}\ ,}
equates such motion with the propagation of an infinite
tower of energy eigenstates $\ket{\nu}$, just like
Eqs.~\schulmanid-\bracketdef\ for $SU(2)$.
 The Yukawa coupling analogous to \PVAdef\ is then
\eqn\newtoyuk{\hf\gbare\,f\,{\partial w\over\partial z'}
\,e^{-i\hat\theta(t')}\,\delta^2(\bx')\ +\ \hbox{H.c.},}
the primed space-time variables referring to the center-of-mass
frame of the moving baryon as in Sec.~2.
Note that Eq.~\newtoyuk\ is a \it special case \rm of Eq.~\toyuk,
with all the $g_\nu$'s equated to a single underlying Yukawa
constant $\gbare$ (to see this, copy the steps in Eq.~\Ddef).
This feature too is just like the higher-dimensional example:
recall the proportionality rule \bf (v) \rm reviewed in Sec.~2.1,
and embodied in the matrix elements \Ddef.

\def\wcl{w_{\rm cl}}
\def\wstat{w_{\rm stat}}
We  have assembled all the ingredients necessary to write down the classical
Euler-Lagrange equation for the meson cloud. The solution,
$\wcl,$ automatically
sums up all contributions to one-meson absorption or emission
from the baryon source, to leading order in the semiclassical
expansion. In solving for $\wcl,$ it is convenient as always
to boost from the Lab
frame to the body-fixed frame of the translating, (iso)rotating,
Lorentz-contracting baryon:
\eqn\wcldef{\wcl(\bx,t)\ =\ e^{i\theta(t')}\,\wstat(\bx')\ .}
In turn, the static meson cloud
$\wstat(\bx)$ finds a solution in the hedgehog ansatz,
composed from \hedgethreeD{b} and \confvar:
\eqn\wstatdef{\wstat(\bx)\ =\ {z\over r}\,\tan{F(r)\over2}\ ,
\quad r=\sqrt{z\bar z}\ .}
Finally, the profile $F(r)$ obeys
\eqn\Ftoy{\eqalign{F''+{1\over r}F'-{1\over2r^2}\sin2F\ &=\
{\gbare(\lam)\over f}\,
\sec^2{F(r)\over2}\,{\partial\over\partial r}\deltalam(r)
\cr &=\ -{\lam^2\gbare(\lam)\over\pi f}\,
\sec^2{F(\laminv)\over2}\,\delta(r-\laminv)\ .}}
This follows straightforwardly from \HamhedgethreeD{b},
 \newtoyuk\ and \wstatdef,
plus an obvious transcription of the $\delta$-function regulator
\mansmear\ to two dimensions.

After this long build-up, we remind the reader of the original motivation
behind this toy model: by recasting the nonlinear $\sigma$
model in two spatial dimensions, we have side-stepped
 Derrick's theorem, and increased
the likelihood of finding a nontrivial UV fixed point to the large-$N_c$
renormalization group equations. We will verify this presently. But
already our efforts have yielded a bonus: unlike Eq.~\homoNLSM, the homogeneous
variant of Eq.~\Ftoy, namely
\eqn\homovar{F''+{1\over r}F'-{1\over2r^2}\sin2F\ =\ 0\ ,}
can be solved analytically. Indeed, switching independent
variables to $\log r$ transforms this into the sine-Gordon
equation, the solutions to which
are the ``baby (anti)Skyrmions''\foot{These
(anti-) solitons in 2+1 dimensions are also the well-known $O(3)$
(anti-) instantons in two Euclidean dimensions \Othree.}
\def\muI{\mu_{\sst\rm I}}
\def\muII{\mu_{\sst\rm II}}
\def\Tan{{\rm Tan}}
\eqn\Ftoysolved{F(r)\ =\ 2\,\Tan^{-1}\muI r\quad\hbox{or}\quad
\pi-2\,\Tan^{-1}\muII r}
where $\muI$ and $\muII$ are arbitrary scale constants.

Figure 10 displays the patched-together solution to Eq.~\Ftoy. In region
II ($r>\laminv$) we choose the solution that decays to zero,
$\FII(r)=\pi-2\,\Tan^{-1}\muII r,$ and fix $\muII$ by normalizing the
large-$r$ falloff to $\gren/(2\pi fr):$
\eqn\muIIdef{\muII\ =\ {4\pi f\over\gren}\ .}
As before, $\gren$ is the renormalized Yukawa constant of the
model.\foot{The ``$N_c$'' scalings are $\gren\sim f\sim\sqn$ in analogy with
the 3+1 dimensional case. Although in this model $N_c$ is no longer identified
with the number of colors, it still usefully parametrizes the semiclassical
expansion. To get the factors of $\pi$ etc.~right in the definition
of $\gren,$ it suffices to solve with a Green's function (cf.~Sec.~4)
the linearized version of \Ftoy, $F''+F'/r-F/r^2=(\gren/f)\deltalam'(r),$
appropriate to the weak-field regime $F\ll1.$ Alternatively, one can
extract the LSZ residue of the meson-mass-shell pole as per Ref.~\DHM.}
Next, pick $\FI(r)=2\,\Tan^{-1}\muI r$ in region I ($r<\laminv$).
Matching $\FI$ to $\FII$ at $r=\laminv$  implies
\eqn\muIdef{\muI\ =\ {\lam^2\gren\over4\pi f}\ .}
The running of $\gbare$ is instantly read off from the slope discontinuity
in Eq.~\Ftoy:
\eqn\gbarerun{\gbare(\lam)\ =\ {\pi f\over\lam^2}\,\cos^2{F(\laminv)\over2}
\,\Big(\FI'(\laminv)-\FII'(\laminv)\Big)\ =\ {\gren\over{\Big[1+
\big({\lam\gren\over4\pi f}\big)^2\,\Big]}^2}\ .}
As anticipated, $\gbare(\lam)$ approaches $\gren$ in the infrared, while
vanishing rapidly ($\sim\,\lam^{-4}$) in the ultraviolet.

Note that for any \it finite \rm value of the cutoff, baryon number in
this model is measured in the mundane way, by the fermion number
operator. Furthermore, the winding number
\windingthreeD{b} of the patched-together meson cloud is zero, since
$F(0)=F(\infty)=0.$ But at infinite $\lam$ the picture looks
very different: the baryons have entirely decoupled from the mesons
thanks to \gbarerun, while the envelope of the sequence of meson clouds
with increasing $\lam$ is obviously a soliton
or Skyrmion with winding number \it unity\rm, namely
$F(r)\equiv\FII(r)$. In this sense, baryon number
can be said to have \it transmuted \rm to winding number at infinite
$\lam.$ And in sharp contrast to the previous Section,
the UV fixed point of our toy ``chiral bag'' model exists and is
nontrivial: a ``baby Skyrmion'' model.

As a more sophisticated alternative,
one may choose to define baryon number density as the \it sum \rm of the usual
explicit fermion number density and the winding number density, which
is more in the spirit of Refs.~\refs{\RGB{--}\GoldJaf}.
 For small $\lam,$ winding number density is negligible
everywhere; the mundane definition is recaptured, and is
entirely concentrated inside the bag, $r\le\laminv$. But for large $\lam$ the
situation is reversed:
explicit fermion number density is screened by negative winding number
density
inside the bag, and the bulk of the baryon number is carried, in the form
of  winding number, by the meson cloud outside the bag. In the
strict continuum limit, the bag is gone altogether,
 and only winding number remains. This is a greatly
simplified variant (with nonrelativistic nucleons,
rather than spectrally-flowing valence and sea quarks) of the
scenario put forward by Goldstone and Jaffe \GoldJaf.

We now exit toyland,
return to 3+1 dimensions, augment the nonlinear $\sigma$ model
\NLSMdef\ by the well-known ``Skyrme term'' to overcome Derrick's theorem,
and examine under what circumstances the statements of the
preceding paragraph can, or cannot, be made.

\newsec{The Skyrme Lagrangian, with the Adkins-Nappi-Witten value of
 $\gpnnren$}

Finally, and most physically, we take for the pion piece of the action
the massless Skyrme Lagrangian \refs{\Skyrme,\ANW}:
\eqn\Lskyrmedef{\calLmeson\ \equiv\ \calLskyrme\ =
{\fpi^2\over16}\,\Tr\,\partial_\mu U^\dagger\partial^\mu U\ +\
{1\over32\es^2}\,\Tr\big[U^\dagger\partial_\mu U\,,\,U^\dagger\partial_\nu U
\big]^2\ .}
We continue to assume that the pions are coupled
to the $I=J$ tower of explicit baryon fields through Eq.~\PVAdef.
Since the two terms in \Lskyrmedef\ scale oppositely under dilatations
\rescaling, Derrick's theorem is avoided, and $\calLskyrme$ supports
a soliton: the original hedgehog Skyrmion (Fig.~11).
In their Skyrme-model treatment, Adkins, Nappi and
Witten take $\fpi=129\,$MeV (vs.~186$\,$MeV experimentally) and
$\es=5.45$ in order to fit the nucleon and $\Delta$ masses \ANW. In the
present \it non\rm-Skyrme-model approach, with explicit nucleons
and Feynman diagrams rather than topology, our above-stated philosophy suggests
instead that we peg these parameters to their experimental values (although
$\es$ is not very well determined by $\pi\pi$ scattering). However,
since our present aims are formal rather than phenomenological,
it is actually
best to leave them unspecified. What we \it do \rm care about are
their $N_c$ assignments:
\eqn\Ncassign{\fpi^2\ \sim\ {1\over\es^2}\ \sim\ N_c\ .}
Thus $N_c/\hbar$ factors out of the action, which justifies our usual
semiclassical manipulations.

Given that the Lagrangian \Lskyrmedef, like the toy model
\newNLSM{b}, supports a soliton, what  might we guess
about the large-$N_c$ renormalization group? Reasoning by analogy
with the preceding Section,
we might expect $\gpnnbare(\lam)$ to vanish in the continuum
limit, with the Skyrmion emerging as the UV fixed point of the family
of meson-baryon ``chiral bag'' models. \it But this naive
scenario cannot generally
be right! \rm

To see why not, let us return to a discussion from Sec.~1.4, and
think about what does---and does not---flow
in our program. What flows are the bare baryon
mass and hyperfine mass splitting parameters $\Mbare(\lam)$ and $\Ibare(\lam),$
as well as the bare Yukawa couplings $\gpnnbare(\lam),$ $g^\bare_{\rho
\sst NN}(\lam),$ etc. What does \it not \rm flow are the purely mesonic
Lagrangian couplings ($\fpi$ and $\es$ in Eq.~\Lskyrmedef, $f$ in
Eq.~\newNLSM{b}, and so forth; see Fig.~12 for a discussion),
 as well as the \it renormalized
\rm parameters $\Mren,$ $\Iren,$ $\gpnnren,$ etc.
These latter quantities, then, are  independent variables at our disposal
in the effective Lagrangian approach---and remain so
even at the supposed endpoint
of the large-$N_c$ renormalization group flow. Now contrast this to
the Skyrmion approach. Assume that $\fpi$ and $\es$ have been specified
once and for all, and look again at the resulting Skyrmion,
Fig.~11. Notice that $\gpnnren,$ rather than being an
additional  tuneable parameter as we have just argued, is instead fixed
by Eq.~\Fasymchiral\  in terms of these meson parameters, as originally
shown by Adkins, Nappi and Witten \ANW:
\def\ganw{g^{\sst\rm ANW}_{\pi\sst NN}}
\eqn\ganwdef{\gpnnren\ \equiv\ \ganw\ \cong\ {18.0\over\es^2\fpi} \ .}
The puzzle can now be stated very clearly: If the Skyrme model is in
fact connected to an effective Lagrangian model by RG flow, then
where, why and how has  the supposedly tuneable Yukawa degree of freedom
in the latter
disappeared?\foot{This paradox never came up in the toy model of Sec.~6
because of \it scale invariance\rm: the Skyrmion $\pi-2\,\Tan^{-1}
\big(4\pi fr/\gren\big)$ exists for any independently-chosen
values of $f$ and $\gren$.}

\def\lamcrit{\lam_{\rm crit}}
The complete
resolution of this paradox is the topic of this Section and the next,
and goes as follows. When $\gpnnren$ is tuned to $\ganw$ precisely,
 then the Skyrme model does indeed  emerge as
the UV fixed point of the ``chiral bag'' models, just as in
 Sec.~6. But for \it all other \rm choices of $\gpnnren,$
the physics is closer to that of Sec.~5: the large-$N_c$ renormalization
group only makes sense up to a critical value of the cutoff $\lamcrit,$
and admits no continuum limit---neither the Skyrme model nor anything
else.

We now flesh this out explicitly. In the
hedgehog ansatz \hedgehog, the  Euler-Lagrange equation \stateuler\
implied by  \Lskyrmedef\ reads
\eqn\skyrmebag{\eqalign{\Big(\,1+{8\sin^2F\over\es^2\fpi^2r^2}\,\Big)F''
&+{2\over r}F'-{\sin2F\over r^2}\,\Big(\,1+{4\sin^2F\over\es^2\fpi^2r^2}
-{4{F'}^2\over\es^2\fpi^2}\,\Big)
\cr&
=\ -{9\lam^3\gpnnbare(\lam)\over2\pi\fpi}\,\delta(r-\laminv)\ .}}
We have again adopted the regulator \mansmear. As above,
we construct solutions $\FI(r)$ and $\FII(r)$
to the homogeneous variant
\eqn\homoskyrme{\Big(\,1+{8\sin^2F\over\es^2\fpi^2r^2}\,\Big)F''
+{2\over r}F'-{\sin2F\over r^2}\,\Big(\,1+{4\sin^2F\over\es^2\fpi^2r^2}
-{4{F'}^2\over\es^2\fpi^2}\,\Big)
\ =\ 0\ ,}
match them up at  $r=\laminv,$
and extract the running of $\gpnnbare$ from the slope discontinuity:
\def\Fskyrme{F_{\rm skyrme}}
\eqn\skyrmerun{\gpnnbare(\lam)\ =\ {2\pi\fpi\over9\lam^3}\,
\Big(\,1+{8\lam^2\sin^2F(\laminv)\over\es^2\fpi^2}\,\Big)\Big(\FI'(\laminv)-
\FII'(\laminv)\Big)\ .}

For the remainder of this Section, we focus on the ``measure zero''
case when $\gpnnren$ is pegged
to its Adkins-Nappi-Witten value \ganwdef. In that event $\FII(r)
\equiv\Fskyrme(r),$ the Skyrmion profile of Fig.~11. For $\FI(r)$, we
take the family of solutions to Eq.~\homoskyrme\
that start at the origin and have increasing slope as $\lam$ itself
is increased (Fig.~13).

We can now evaluate Eq.~\skyrmerun. For small $r$, the Skyrmion profile
may be Taylor-expanded:
\eqn\Ftaylor{\Fskyrme(r)\ =\ \pi-c_1\es\fpi r+{\cal O}(\es\fpi r)^3\ .}
Therefore, the first term in parentheses in \skyrmerun\ is
$1+8c_1^2+{\cal O}(\lam^{-2})$
 in the ultraviolet, where the numerical slope $c_1$ may be
read off Fig.~11.
The second term in parentheses is dominated  for large $\lam$ by
\eqn\FIprime{\FI'(\laminv)\ \sim\ {c_2\lam^2\over\es\fpi}\ ,}
where $c_2$ is another numerical constant.\foot{This $\lam^2$ behavior
is surprising, since generically along curve I, $\FI'(r)\sim\lam$ instead,
as is easily argued. However, very near the point where $\FI$ crosses
$\pi$, its slope changes over to a $\lam^2$ behavior. An interesting way
of proving this statement (which we first observed numerically) is
with the calculus of variations. Specifically, if one performs a
small deformation of the cloud analogous
to that shown in Fig.~15 below, and demands that
the first variation vanish (as it must), one derives Eq.~\FIprime, including
an explicit expression for $c_2$ in terms of $c_1$ and a definite
 integral of the  Skyrme Hamiltonian.}
Combining these expressions then gives
\eqn\gbareskyrme{\gpnnbare(\lam)\
\matrix{& {\sst\lam\rightarrow\infty}\cr&\propto\cr&{}}\ {1\over\es\lam}\ .}
Thus, just as  in the toy model, the baryons  decouple from the pions
in the ultraviolet limit, at which point ordinary
 baryon number $\sum_{i_z,s_z}\big(N^\dagger N+\Delta^\dagger\Delta+\cdots
\big)$ effectively transmutes into the winding number of the
pion cloud, Eq.~\windingthreeD{a}, in the manner discussed at the end
of Sec.~6.

\newsec{The Skyrme Lagrangian, with an incorrect choice of $\gpnnren$}

Finally, we analyze the ``generic'' case when $\gpnnren$ differs from
its Skyrme-model value $\ganw.$ In what way is the solution to
Eqs.~\skyrmebag-\skyrmerun\ affected? Consider how, in practice,
one constructs $\FII(r)$ numerically. One first sets $\FII\,\sim\,
3\gpnnren/(2\pi\fpi r^2)$ in the asymptotic regime $r\gg(\es\fpi)^{-1},$
then  integrates inwards towards the origin, and in so doing,
encounters a surprise:
 rather than intercepting the $y$ axis at $\pi$ as in the
previous case with $\gpnnren=\ganw,$ or diverging for small $r$ as in the
nonlinear $\sigma$ model of Sec.~5, $\FII$ invariably hits the $y$
axis at a \it half\rm-integral multiple $(n+\hf)\pi,$ as shown in Fig.~14a.
This behavior can be confirmed analytically. Linearizing Eq.~\homoskyrme\
about such half-integral values forces
\eqn\FIIUV{\eqalign{\FII(r)
\ \matrix{&{\sst r\rightarrow0}\cr&\longrightarrow\cr&{}}\ \
&\big(n+\hf\big)\pi\ +\ \hbox{const.}\times r^{(1+i\sqrt{3})/2}
\ +\ \hbox{const.}\times r^{(1-i\sqrt{3})/2}
\cr&
\qquad +\ {\cal O}(r^{3/2})
\cr=\
&\big(n+\hf\big)\pi\ -\ (\mu_1r)^{1/2}\cos\big(\textstyle{\sqrt{3}\over2}
\log\mu_2r\big)\ +\ {\cal O}(r^{3/2})\ .}}
The fact that there are two independent constants here,
 $\mu_1$ and $\mu_2$, shows that this is indeed a  generic family of
solutions to Eq.~\homoskyrme.\foot{The only other
self-consistent solutions at small $r$ start at integer multiples of $\pi,$
and are of the form $n\pi-\mu r+{\cal O}(r^3),$ e.g., the Skyrmion itself.
Unlike Eq.~\FIIUV, these are only \it one\rm-parameter families (parametrized
by $\mu$), so one never ``accidentally'' stumbles upon them when numerically
integrating inwards. Amusingly, the same phenomenon holds true at \it
large \rm $r$ as well \Syracuse\ (cf. Fig.~9b):
 there are generic two-parameter
families asymptoting to $(n+\hf)\pi,$ and special one-parameter families
(again, including the Skyrmion) terminating at $n\pi.$
So the 3+1 dimensional Skyrmion is ``special''  both at $r=0$ and
at $r=\infty,$ hence
has no free parameters. Of course, finiteness of the energy
requires integer, not half-integer, boundary conditions at both ends,
which we enforce on our patched-together ``chiral bags.''}
 The bizarre oscillatory behavior of this essential
singularity is verified in Fig.~14b.

Of course, the
family of $\FI$ curves is still given by Fig.~13, and $\gpnnbare(\lam)$
is again read off from the slope discontinuity \skyrmerun\ at
 $r=\laminv,$ so one can still patch together
 a \it bona fide \rm solution to Eq.~\skyrmebag. But, as always,
 this solution is
only physically relevant if it is \it locally stable \rm against small
deformations of the meson cloud.

We have carefully
investigated this issue of local stability (albeit only within
the hedgehog ansatz), and identified one
dangerous mode, described in Fig.~15. Since we are perturbing about
an Euler-Lagrange solution, first variations necessarily vanish. Instead,
our stability analysis focuses on the sign of the coefficient $Q$
of quadratic variations in this mode, a positive (negative) value
indicating (in)stability. We have calculated (see Appendix A for details):
\eqn\Qdef{Q\ \matrix{&{\sst \lam\rightarrow\infty}\cr&\longrightarrow\cr&{}}
 \ \hbox{const.}\times\big(\es^{-3}\fpi\lam\big)^{1/2}
\cos\textstyle\big({\sqrt{3}\over2}\log\mu_2\laminv\big)\
+\ {\cal O}(\lam^0)\ .}
Therefore, for
fixed $\gpnnren\neq\ganw,$ and sufficiently large $\lam,$ the model exhibits
 alternating
``phases'' of local stability and instability along a large-$N_c$
renormalization group trajectory. Mathematically,
this behavior is traceable to the
short-distance essential singularities \FIIUV\ in the
solutions to Eq.~\homoskyrme. Each such phase lasts
half a period of the sinusoidal oscillation, thus:
\eqna\stab
$$\eqalignno{\hbox{\bf unstable phases}&:\cr
\quad
\lam\in[\lcrit\,,\,\kappa\lcrit]\ &,\
\lam\in[\kappa^2\lcrit\,,\,\kappa^3\lcrit]\ ,\
\lam\in[\kappa^4\lcrit\,,\,\kappa^5\lcrit]\ ,\ \cdots
&\stab a
\cr
{\hbox{\bf stable phases}}&:\cr
\quad
\lam\in[\kappa\lcrit\,,\,\kappa^2\lcrit]\ &,\
\lam\in[\kappa^3\lcrit\,,\,\kappa^4\lcrit]\ ,\
\lam\in[\kappa^5\lcrit\,,\,\kappa^6\lcrit]\ ,\ \cdots\ .
&\stab b
}$$
Here $\kappa=\exp(2\pi/\sqrt{3})$ is the constant whose logarithm
equals half a period, and $\lcrit$ is defined as the
first appearance of this instability as the large-$N_c$ renormalization
group is pushed into the ultraviolet. $\lcrit$ is analogous to the
Landau pole in quantum electrodynamics: it is the scale beyond which
one cannot push the cutoff while requiring that the theory be stable.
Of course, the value of $\lcrit$ depends sensitively on $\gpnnren$; in
general, the closer the latter is to $\ganw,$ the greater we expect
$\lcrit$ to be. In this language, the results of the previous Section
may be understood as the statement that $\lcrit\rightarrow\infty$
(stability regained) as $\gpnnren\rightarrow\ganw.$

As in the nonlinear $\sigma$ model of
Sec.~5, one can also patch together pion clouds
with nonzero winding number, but each of these is again locally stable only
for finite ranges in $\lam$ (and potentially locally \it un\rm stable
against small deformations \it away \rm from the hedgehog ansatz).
While the nature of the mathematical obstruction is somewhat different,
the conclusion here is the same as
for Sec.~5: the Skyrme Lagrangian, coupled to the baryon
tower, admits no continuum limit when $\gpnnren$ differs
from its canonical Skyrme-model value $\ganw.$ In this interesting way
the ``large-$N_c$ renormalization group as filter'' idea is realized.
\vskip.15in

We thank Jim Friar,
Jim Hughes, Alex Kovner, Aneesh Manohar and Dick Silbar for useful
comments, and Jessica Binder for help with the artwork.
 MPM is indebted to the Swansea physics department for
hospitality during the time that much of this work was carried out.
This draft was posted on the occasion of ND's 30th birthday.

\appendix{A}{Cloud collapse in the Skyrme example when $\gpnnren\neq\ganw.$}

\leftline{\it **Derivation of the local instability**}

\def\Hbag{H_{\rm tot}}
The primary goal of this Appendix is to derive the oscillatory
expression \Qdef\ for
the Gaussian coefficient $Q$ of the mode shown in Fig.~15. For this
purpose, we need to construct
the total energy of the cloud, $\Hbag[F],$ keeping only the
terms of ${\cal O}(N_c)$. The usual
Skyrme-model Hamiltonian density is the sum of the contributions from the
 2-derivative and 4-derivative terms,
\def\calHskyrme{{\cal H}_{\rm skyrme}}
\def\calHtwoderiv{{\cal H}_{2\hbox{-}\rm deriv}}
\def\calHfourderiv{{\cal H}_{4\hbox{-}\rm deriv}}
\eqn\Hskyrmedef{\eqalign{\calHskyrme\ &=\ \calHtwoderiv\ +\ \calHfourderiv\cr&
=\ {\fpi^2\over8}\Big({F'}^2+2{\sin^2F\over r^2}\Big)\ +\
{1\over2\es^2}\Big({\sin^4F\over r^4}+2{{F'}^2\sin^2F\over r^2}\Big)\ .}}
$\Hbag$ is itself the sum of three distinct parts, the contributions
from region I, region II, plus the negative-definite
energy of the Yukawa interaction with the baryonic source:
\def\HI{H_{\sst\rm I}}
\def\HII{H_{\sst \rm II}}
\def\Hyuk{H_{\rm yukawa}}
\eqn\Hbagdef{\Hbag[F]\ =\ \HI[F]+\HII[F]+\Hyuk[F]}
where
\eqn\HIdef{\HI[F]\ =\ 4\pi\int_0^{\laminv} r^2dr\,\calHskyrme\ ,\quad
\HII[F]\ =\ 4\pi\int_{\laminv}^\infty r^2dr\,\calHskyrme\ ,}
and
\eqn\Hyukdef{\Hyuk[F]\ =\ -{\textstyle{9\over2}}\fpi\lam\gpnnbare
(\lam) F(\laminv)}
The point-like form of the latter is due to our special choice of
smearing, Eq.~\mansmear.
The defining equation for $F$, Eq.~\skyrmebag, is precisely the Euler-Lagrange
equation
\eqn\bageuler{{\delta\over\delta F}\,\Hbag[F]\ =\ 0\ .}

\def\arg{\big((1-\eta)\laminv\big)}
Suppose that $F(r)$ is a patched-together solution to Eq.~\skyrmebag,
and consider the effect of the small deformation
 of Fig.~15 on each of the three
parts in Eq.~\Hbagdef. As constructed, the deformation keeps the matching
point at $r=\laminv$ always, but changes the value of $F$ at this point
from $\FI(\laminv)$ to $\FI\arg$. With this sign convention, a positive
value of $\eta$ lowers the patched-together cloud so that it is closer
to the vacuum $F(r)\equiv0,$ whereas a negative value of $\eta$ raises
the curve so that it is closer to the Skyrmion of Fig.~11. We will treat
$\eta$ as an infinitesimal parameter, and will find it fruitful to keep
all terms through order $\eta^3.$ The effect on $\Hyuk$ is
immediate:\foot{From now on, $\FI$ is short for $\FI(\laminv)$ and likewise
for its derivatives. Note that this equation is a perturbative expansion
in $\eta,$ \it not \rm in $\laminv$: generically along the solution curve
in region I, $\FI'(r)\sim{\cal O}(\lam),$
$\FI''(r)\sim{\cal O}(\lam^2),$ $\FI'''(r)\sim{\cal O}(\lam^3),$
and so forth, so that the terms in brackets are each nominally ${\cal O}
(\lam^0).$ (Two important exceptions to this generic behavior along
curve I: when $\FI(r)$ is near $\pi$ then $\FI'(r)\sim\lam^2$ rather
than $\lam$ as explained earlier, and also, when $\FI(r)$ is near
$\pi/2$ then $\FII''(r)\sim\lam^{3/2}$ rather than $\lam^2$ as implied by
%danger
 Eq.~(A.7).) Another possibly confusing point:
$\gpnnbare(\lam)$ does \it not \rm vary under
this deformation; it is just a Lagrangian parameter like any other for the
purposes of the variational calculus.}
\eqn\pertHyuk{\Hyuk \longrightarrow
-{\textstyle{9\over2}}\fpi\lam\gpnnbare(\lam)\Big[\FI-\eta\laminv\FI'
+{1\over2!}\eta^2\lam^{-2}\FI''-{1\over3!}\eta^3\lam^{-3}\FI'''+{\cal O}
(\eta^4)\Big] .}

Next, we examine $\HI.$
By design, $\FI(r)$ is supposed to remain a solution to Eq.~\homoskyrme\
as we perform the deformation.
For large $\lam$ (and therefore, small $r$),
the important terms in this equation are
those that derive from varying $\calHfourderiv,$ namely:
\eqn\fourderiveqn{{8\sin^2\FI(r)\over r^2}\,\FI''(r)
-{\sin2\FI(r)\over r^2}\Big({4\sin^2\FI(r)\over r^2}
-4\FI'(r)^2\,\Big)\ \approx\ 0\ .}
The key observation here is that this (approximate) equation is \it scale
invariant\rm, so that the solution in region I which satisfies
\def\new{{\rm new}}
 the new perturbed boundary conditions,
$\FI^\new(0)=0$ and $\FI^\new(\laminv)=\FI\arg$, is just
$\FI^\new(r)\approx\FI\big((1-\eta)r\big)$ for all $r$ in this
range. A straightforward
Taylor expansion then yields
\eqn\pertHI{\eqalign{\HI\ \longrightarrow\ 4\pi&\int_0^{\laminv} r^2dr\,
\calHfourderiv\big[\FI\big((1-\eta)r\big)\,\big]
\cr
=\ 4\pi(1-\eta)&\int_0^{(1-\eta)\laminv} \tilde r^2d\tilde r\,
\calHfourderiv\big[\FI(\tilde r)\,\big]
\cr=\
4\pi(1-\eta)\Bigg\{&\int_0^{\laminv} \tilde r^2d\tilde r\,
\calHfourderiv\big[\FI(\tilde r)\,\big]
\ -\ {\eta\laminv\over2\es^2}\,\Big(2{\FI'}^2\sin^2\FI+\lam^2\sin^4
\FI\Big)
\cr
+\ {1\over2!}&{\eta^2\lam^{-2}\over2\es^2}\,\Big(
2{\FI'}^3\sin2\FI+4\FI'\FI''\sin^2\FI
\cr&
+2\lam^2\FI'\sin2\FI\sin^2\FI
-2\lam^3\sin^4\FI\Big)
\cr
 -\ {1\over3!}&{\eta^3\lam^{-3}\over2\es^2}\,\Big(
{\FI'}^4(4-8\sin^2\FI)+10{\FI'}^2\FI''\sin2\FI
\cr&
+4(\FI'\FI'''+{\FI''}^2)\sin^2\FI
+\lam^2{\FI'}^2(12\sin^2\FI-16\sin^4\FI)
\cr&
+(2\lam^2\FI''-8\lam^3\FI')\sin2\FI\sin^2\FI+6\lam^4\sin^4\FI
\Big)
\ +\ {\cal O}(\eta^4)\Bigg\}}}
In the first equality we have changed integration variables to
 $\tilde r=(1-\eta)r,$ and  exploited the scale-covariance of
$\calHfourderiv$.

This cumbersome expression simplifies considerably if one discards
all terms of order $\lam^0,$ keeping only those of order $\lam$ and
$\lam^{1/2}.$ At the point $r=\laminv$ where $\FI=\FII,$
the relevant scalings are
\eqn\scalinglist{\eqalign{&\hf\sin2\FI\approx(n+\hf)\pi-\FI\sim
\lam^{-1/2}\ ,\quad\sin^2\FI\approx1\ ,
\cr&\FII'\sim\lam^{1/2}\ ,\quad\FI'\sim\lam\ ,\quad
\FI''\sim\lam^{3/2}\ ,\quad\FI'''\sim\lam^3}}
thanks to Eqs.~\FIIUV\ and \fourderiveqn. Equation \pertHI\ collapses to
\eqn\newHI{\eqalign{\HI\ =\
4\pi(1-\eta)\Bigg\{&\int_0^{\laminv} \tilde r^2d\tilde r\,
\calHfourderiv\big[\FI(\tilde r)\,\big]
\ -\ {\eta\laminv\over2\es^2}\,\Big(2{\FI'}^2+\lam^2\Big)
\cr
+\ {1\over2!}&{\eta^2\lam^{-2}\over2\es^2}\,\Big(
2{\FI'}^3\sin2\FI+4\FI'\FI''+2\lam^2\FI'\sin2\FI-2\lam^3\Big)
\cr
 -\ {1\over3!}&{\eta^3\lam^{-3}\over2\es^2}\,\Big(
-4{\FI'}^4+4\FI'\FI'''-4\lam^2{\FI'}^2
-8\lam^3\FI'\sin2\FI+6\lam^4\Big)
\cr&
\qquad\qquad\qquad
\ +\ {\cal O}(\eta^4)\ +\ {\cal O}(\lam^0)\Bigg\}\ .}}

\def\rmax{{R_{\rm max}}}
Finally we consider region II, for which the deformation at short
distances consists of adding the constant $\FI\big((1-\eta)\laminv\big)
-\FI(\laminv)$ to the right-hand side of Eq.~\FIIUV.
The dominant contribution to the energy in this region
comes from the $\sin^4\FII(r)/(2\es^2r^4)$ piece
of Eq.~\Hskyrmedef, which is the only term whose integral diverges
as $\lam\rightarrow\infty$. Another short exercise in Taylor expansion
gives\foot{The appearance of $\FII'$ on the right-hand side is
due to the observation that $\int_{\laminv}^\rmax{dr\over r^2}\,
r^\lambda\,=\,{d\over dr}\,r^\lambda{\big|}_{r=\laminv}+
{\cal O}(\Lambda^0)$ precisely
when $\lambda=(1\pm i\sqrt{3})/2$ as per Eq.~\FIIUV.}
\eqn\pertHII{\eqalign{&\HII\ \longrightarrow\ 4\pi\int_{\laminv}^\rmax
{r^2\over2\es^2r^4}\,dr\,
\sin^4\big[\FII(r)+\FI\big((1-\eta)\laminv\big)-\FI(\laminv)\big]
\cr&=\ -{2\pi\over\es^2}\Big[-\lam-4\eta\laminv\FI'\FII'
+2\eta^2\lam^{-1}{\FI'}^2
\cr&\qquad\qquad-{\eta^3\lam^{-3}\over3!}\big(12\lam\FI'\FI''+4\FII'\FI'''
-4\FII'{\FI'}^3\big)\ \Big]\ +\ {\cal O}(\eta^4)\ +\ {\cal O}(\lam^0)\ .}}
The details of the upper limit of integration $\rmax$ are unimportant, as they
do not conceivably involve divergent terms in $\lam$.

Assembling Eqs.~\pertHyuk, \newHI\ and \pertHII\ yields the series expansion
in the small-deformations parameter:
\eqn\bagseries{\Hbag(\eta)\ =\ Z\ +\ \eta L\ +\ \eta^2Q\ +\ \eta^3C
\ +\ {\cal O}(\eta^4)\ .}
Focusing first on the quantity of greatest interest, the quadratic
coefficient $Q$, we find
\eqn\Qanswer{Q\ =\ {2\pi\over\es^2}\Big(\lam^{-2}{\FI'}^3+
\FI'\,\Big)\,\sin2\FI\ +\ {\cal O}(\lam^0)\ ,}
using Eq.~\skyrmerun\ to eliminate $\gpnnbare$. Notice that $Q$ is only
${\cal O}\big(\lam^{1/2}\big),$ the ${\cal O}(\lam)$ pieces having
canceled between \newHI\ and \pertHII. This confirms Eq.~\Qdef.

The energy of this patched-together cloud is given by the zeroth-order
coefficient $Z$:
\eqn\Zanswer{\eqalign{Z\ &=\
4\pi\int_0^{\laminv} \tilde r^2d\tilde r\,\calHfourderiv\big[\FI(\tilde r)\,
\big] \ +\ {2\pi\over\es^2}\Big(\lam-4\FI\FI'+4\FI\FII'\Big)
\cr&=\ {2\pi\over\es^2}\Big(2\laminv{\FI'}^2-4\FI\FI'+4\FI\FII'\Big)\
+\ {\cal O}(\lam^{0}).}}
To eliminate the integral in the second equality, we have used the
fact that since the cloud is a solution to Eq.~\bageuler\ at $\eta=0,$
the linear term $L$ must vanish:
\eqn\lincancel{0\ =\ L\ = {2\pi\over\es^2\lam}
\big(2{\FI'}^2-\lam^2\big)\ -\
4\pi\int_0^{\laminv} \tilde r^2d\tilde r\,
\calHfourderiv\big[\FI(\tilde r)\,\big] \ +\ {\cal O}(\lam^{0}).}
Numerically, $Z$ turns out to be \it negative definite \rm for the important
case $n=0$ in Eq.~\FIIUV, corresponding to the coupling range
$0<\gpnnren<\ganw,$ as $Z$ is then
dominated by the attractive Yukawa contribution \Hyukdef.

\leftline{\it **A nontrivial consistency check**}

Finally, the cubic coefficient is
\eqn\Canswer{C\ =\ {4\pi\over3\es^2\lam^3}
\Big({\FI'}^4+\lam^2{\FI'}^2\,\Big)\ +\ {\cal O}(\lam^{1/2})\ .}
The reason $C$ is interesting is that there is a stringent
consistency check on our calculations in terms of the ratio
$R=Q/C$, which reads
\eqn\consistency{\FI'{\partial R\over\partial \FI}
+\FI''{\partial R\over\partial \FI'} +\FI'''{\partial R\over\partial \FI''}
+\cdots\ =\ -3\lam+{\cal O}(\lam^{1/2})\ .}
Thanks to Eq.~\scalinglist, this is indeed satisfied by Eqs.~\Qanswer\ and
\Canswer.

To understand the source of this consistency check, look at Fig.~16.
It shows that for every patched-together solution of the
type we have been discussing, there must be another nearby solution,
labeled ``A'' and ``B'', respectively.  Algebraically,
this follows from the
cubic polynomial \bagseries, which can be rewritten more explicitly:
\eqn\newseries{\eqalign{\Hbag(\eta,\lam)\ \cong\ &Z\big(
\FI(\laminv)\,,\,\FI'(\laminv)\,,\,\cdots\big)
\ +\ \eta^2Q\big(\FI(\laminv)\,,\,\FI'(\laminv)\,,\,\cdots\big)\cr
\ +\ &\eta^3C\big(\FI(\laminv)\,,\,\FI'(\laminv)\,,\,\cdots\big)\ .}}
Aside from the starting solution ``A,'' which is at $\eta=0$ by definition,
Eq.~\newseries\ implies
 a new solution ``B'' which stationarizes $\Hbag(\eta)$,
at $\eta_*\cong-2R/3\sim\lam^{-1/2}.$ Alternatively, one could
start one's Taylor expansion at ``B,''
\eqn\newestseries{\eqalign{\Hbag(\eta\,,\,\lam)\ \cong\ &Z\big(
\FI((1-\eta_*)
\laminv)\,,\,\FI'((1-\eta_*)\laminv)\,,\,\cdots\big)
\cr \ +\ &
\eta^2Q\big(\FI((1-\eta_*)\laminv)\,,\,\FI'((1-\eta_*)
\laminv)\,,\,\cdots\big)\cr
\ +\ &\eta^3C\big(\FI((1-\eta_*)
\laminv)\,,\,\FI'((1-\eta_*)\laminv)\,,\,\cdots\big)\ ,}}
and demand that the original solution ``A''
be recovered at $\eta=\eta_{**}.$ Consistency between these alternative
starting points forces $1-\eta_{**}=(1-\eta_{*})^{-1}.$ When $\lam\gg1$
so that $\eta_*$ and $\eta_{**}\ll1,$ this in turn implies
$\eta_{**}\cong-\eta_*$, or in other words,
\eqn\Rcondition{
R\big(\FI((1-\eta_*)\laminv)\,,\,\FI'((1-\eta_*)\laminv)\,,\,\cdots\big)
\ \cong\ -R\big(\FI(\laminv)\,,\,\FI'(\laminv)\,,\,\cdots\big)\ ,}
whereupon  Eq.~\consistency\ follows from a Taylor expansion in $\eta_*.$

\leftline{\it **A numerical example**}

Finally we describe a numerical example which bears out the above
picture of ``twinned'' solutions ``A'' and ``B,'' and is further instructive in
answering the question, What does the locally unstable cloud ``A'' collapse
into?

Specifically, we have patched together a solution to Eq.~\skyrmebag\ by fixing
$\gpnnren=.65\ganw,$ and integrating Skyrme's equation for
$\FII(r)$ inwards from infinity, to a matching point at
 $r=\laminv$ where we fix $\lam=6.5\es\fpi.$ The ``region
I'' curve $\FI(r)$ is
then the solution to Skyrme's equation with boundary conditions
$\FI(0)=0$ and $\FI(\laminv)=\FII(\laminv)=.68\pi$. From the slope
discontinuity one finds $\gpnnbare=8.2\es^{-2}\fpi^{-1}=.46\ganw,$
while the total cloud energy \Hbagdef\ is $-183\fpi/\es, $ which is
within a few percent of the large-$\lam$ approximation \Zanswer. Since at
the matching point $F(\laminv)>\pi/2,$ we expect this to correspond to
the locally unstable configuration
``A'' in Fig.~16. Can one then uncover a solution with
$F(\laminv)<\pi/2,$ corresponding
to the nearby locally stable (but still globally unstable) configuration
``B''? Sure enough, for the same Lagrangian parameters  $\lam$
and $\gpnnbare(\lam),$ one finds a second solution matched at
 $F(\laminv)=.41\pi,$ whose renormalized coupling turns out to be
 tiny, $\gpnnren=.062\ganw,$
and whose total cloud energy is $-205\fpi/\es$, a little lower than that of
``A'' as Fig.~16 suggests.

Figure 16 further suggests the existence of
 a \it third \rm solution, labeled
``C,'' which is energetically much more favorable than either
``A'' or ``B.'' We can argue for the necessity of such a solution,
by noticing that for sufficiently negative $\eta,$ eventually the cost
in gradient energy must overwhelm the gain in the Yukawa interaction;
therefore there must be a new minimum for some finite, negative $\eta.$
One naturally guesses that this new solution looks more like the
Skyrmion, meaning that the value of $F(\laminv)$ is closer to
$\pi$ than $\pi/2$, and that the asymptotics of the tail is closer
to Fig.~11. And indeed, for the same Lagrangian parameters,
we have found a solution for which $F(\laminv)=1.18\pi,$
 $\gpnnren=1.08\ganw$, and the total bag energy is $-338\fpi/\es.$

While we believe that, for this fixed choice of $\lam,$
solution ``C'' is both a local and
global minimum of the energy, certainly if it is extrapolated far enough
into the ultraviolet with the large-$N_c$ renormalization group, it
too eventually develops an instability. Although we have not
pushed the numerics, it is tempting to conjecture that the cloud
then collapses
to one whose renormalized coupling is closer still to $\ganw,$ and
that by continuing in this way---alternating RG flow with cloud collapse
to a new renormalized coupling, followed by renewed RG flow to still
higher $\lam,$ and so forth---one iterates one's way to $\ganw$ precisely.
\vfil\eject
\listrefs
\centerline{\bf Figure Captions}
\centerline{\tt Contact  mattis@skyrmion.lanl.gov  for hard copies}
\centerline{\tt if you cannot process the figures from hep-ph}
1. Three types of large-$N_c$ models of the strong interactions, and
the relationships between them. This paper examines two of the
three arrows, Effective QFT
$\Longrightarrow$
``Chiral bags,'' and ``Chiral bags''
$\Longrightarrow$
Skyrmions. The third relation, Skyrmions
$\Longrightarrow$
Effective QFT, is examined in depth in Ref.~\DHM.

2. (a) A bare meson-baryon coupling $\gbare\sim\sqrt{N_c},$
which we shall refer to generically as a ``Yukawa coupling'' even if
it involves derivatives.
Henceforth, directed lines are baryons, undirected lines are mesons.
(b) A simple radiative correction to (a). Since the 3-meson
vertex $\sim\,1/\sqrt{N_c},$ this graph too $\sim\sqrt{N_c}.$ Therefore,
it is a leading-order contribution to $\gren.$
(c) A more complicated contribution to $\gren$ which is likewise
leading-order. The general rule is: the leading-order graphs are the
ones for which, if one erases the baryon line(s), one is left with
meson tree(s).
(d) An example of a subleading contribution to $\gren.$
The purely mesonic loop costs one power of $N_c$, so this graph
$\sim\,1/\sqrt{N_c}.$

3. (a) A ``seagull'' interaction, in which more than one meson
interacts with the baryon at the same space-time point. The ellipses
allow for yet more meson lines than shown meeting at the vertex. (b)
A $Z$-graph, in which the baryon runs backwards in time over
an interval. These arise in the decomposition of Feynman graphs
into time-ordered diagrams. Time runs upwards in this diagram.
(c) An  effective seagull,
or ``$Z$-gull,'' implied by (b). For Yukawa couplings that obey the
$I_t=J_t$ rule, (c) is suppressed by $1/N_c^2$ compared to the
bare seagull shown in (a).
%The vertex is approximated
%by $\gbare^2/2\Mren\sim N_c^0$ up to nonlocal $1/N_c$ corrections.
%(d) A hybrid diagram featuring both bare seagulls and
%backwards-in-time baryons. (e) The effective $Z$-gull obtained
%from (d).
%
%4. (a) A Feynman graph which contributes to the renormalized Yukawa
%coupling, containing not only a tree-level meson branching,
%but also a meson radiative correction to the
%baryon self-energy. All possible time orderings are implicitly
%summed. (b) The ``$Z$-gull'' induced when baryon propagator
%\#2 runs backwards in time and so effectively contracts to
%a point (cf.~Fig.~3). This time-ordered graph contains
%a purely mesonic loop, which costs a factor of $1/N_c,$
%hence it does not contribute in the
%large-$N_c$ limit (cf.~Fig.~2d).  The self-energy
%corrections do not affect this counting as they
%separately exponentiate (cf.~Sec.~3.2 below).
%(c) The ``$Z$-gull'' induced when baryon propagator
%\#1 runs backwards in time. In contrast to (b), here
%there are no purely mesonic
%loops, so this is a  leading-order time-ordered diagram.

4. The complete set of leading-order corrections of the type
shown in Figs.~2a-c. The oval blob is understood to contain
all tree-level meson branchings (no  loops).
 There is an explicit sum over the $n!$
attachments of the blob to the baryon line.

5. The graphical Born-series solution of Eq.~\euler. The line terminating
in a square is our notation for $\vpicl,$ the oval blob
contains all tree-level meson branchings, and the Yukawa source
$\cal Y$ is short for the  right-hand side of Eq.~\euler.

6. An equivalent rewrite of Fig.~5, with the baryon explicitly
drawn (it is implicit in the Yukawa source of Fig.~5). The difference
with Fig.~4 lies solely in the time-ordering up the baryon line
(a $1/N_c$ difference).

7. (a) The left-hand side of Eq.~\cloudaction, formed from
Eqs.~\PVAdef\ and \Lpidef, in the graphical
language of Fig.~5. The third summand stands for the sum of all the vertices
in the potential $V(\vpicl).$ Varying (a) with respect to $\vpicl$ gives
 Eq.~\euler. (b) Born-series rewrite of (a) using
the expansion shown in Fig.~5,  with combinatoric
factors suppressed. (c) The meaning of (b) as
baryon self-energy and meson-baryon vertex corrections, interpreted in the
original Feynman-diagrammatic language of Fig.~2.
These corrections, shown to the right of the baryon line, need to be inserted
in all possible locations up this line, as dictated by Eq.~\unity.

8. Typical hedgehog profile $F(r).$ In the large-$r$ regime, all the
complicated
dependence on the meson potential $V(\vpi)$ and on the form of the
regulator $\deltalam(r)$ has been reduced to the single
parameter $\gpnnren$ which measures the height of the tail.

9. Construction of the patched-together cloud $F(r)=\theta(\laminv-r)\FI(r)
+\theta(r-\laminv)\FII(r)$ for the model of Sec.~5.
(a) The curve $\FII(r),$ which is uniquely specified by the asymptotic form,
Eq.~\Fasymchiral. The $x$ axis is in units of
the ${\cal O}(N_c^0)$ length $(3\gpnnren/2\pi\fpi)^{1/2}.$
$\FII(r)$ blows up like $1/r$ for small $r$.
 (b) The curve $\FI(r)$. The cross indicates the curve's
maximum, $\FI^{\rm max}\cong .58\pi$.  (We have also marked by a cross
in (a) the point where $\FII(r)=\FI^{\rm max}$ which defines the
critical scale $r=\lam_2^{-1}$ discussed in the text.)
 In Region A its slope is positive and in Region B it is negative;
eventually the curve asymptotes to $\pi/2.$ The scale of the $x$ axis is
purposefully not displayed, because this curve stands for the entire
one-parameter \it family \rm of curves related by dilatations:
$\FI(r)\rightarrow\FI(\lambda r)$. For any given value of the cutoff
$\lam,$ this $\lambda$ is to be adjusted so that $\FI(\laminv)=
\FII(\laminv)$.

\def\rtilde{\tilde r}
\def\lamt{\tilde\lam}
10. The patched-together solution to Eq.~\Ftoy, namely: $F(\rtilde)=
2\Tan^{-1}(\lamt^2\rtilde)\theta(\lamt^{-1}-\rtilde)+
(\pi-2\Tan^{-1}\rtilde\,)\theta(\rtilde-\lamt^{-1})$, where the
dimensionless
variables are $\rtilde=4\pi fr/\gren$ and $\lamt=\gren\lam/4\pi f.$
Shown is the curve for $\lamt^{-1}=4.$

11. The original Skyrmion \refs{\Skyrme,\ANW}. The $x$ axis is in
${\cal O}(N_c^0)$ length units $(\es\fpi)^{-1}.$

12. (a) A generic bare 3-meson vertex. (b) and (c): Radiative
corrections to the vertex, both of which are down by $1/N_c$ compared to (a),
thanks to selection rule \bf (i) \rm in Sec.~2.1.
Hence, unlike Yukawa couplings, purely mesonic vertices do \it not \rm run
in the large-$N_c$ renormalization group at leading order.
It is therefore reasonable to fix the bare mesonic parameters
to the experimental data at the outset.

13. Family of curves $\FI(r)$ that solve Skyrme's equation.
 When $\gpnnren\equiv\ganw,$ $\FII(r)$ is just the Skyrmion
profile shown in Fig.~11.

14. (a) A typical curve $\FII(r)$ when $0<\gpnnren<\ganw.$ As explained
in the text, rather than approaching $\pi$ for small $r$, this
curve spirals into $\pi/2.$ The $x$ axis is in units $(\es\fpi)^{-1}.$
(b) The short-distance behavior of (a) in logarithmic variables,
confirming the oscillatory behavior of Eq.~\FIIUV. The variable
plotted on the $x$ axis is $\log(\es\fpi r),$ while the $y$ axis
is $(\es\fpi r)^{-1/2}(\FII-\pi/2)$ as suggested by Eq.~\FIIUV.

15. A dangerous small deformation of the patched-together
pion cloud. For alternating regions in $\lam,$ the cloud energy
proves to be quadratically unstable against this deformation, and the cloud
instantly collapses to another configuration closer to the Skyrmion.
The deformation is defined as follows. Hold the matching point
fixed at $r=\laminv,$ but raise the  value  of $\FI(\laminv)=
\FII(\laminv)$ infinitesimally, while letting the entire curve $\FI(r)$
relax to a new solution of Eq.~\homoskyrme\ for $r<\laminv$.
For $r\ {>\atop\sim}\ \laminv,$ raise $\FII(r)$ by a constant
amount so that the curves stay matched,
 then merge this curve smoothly to an $r^{-2}$ falloff at large
distances (the details of this merging are irrelevant for
sufficiently large $\lam$).

16. Total cloud energy $\Hbag$ as a function of the
matching-point value $\FI(\laminv)=\FII(\laminv).$ For large $\lam,$ the
existence of nearby solutions ``A'' and ``B'' follows from
the cubic polynomial, Eq.~\bagseries, remembering that
$L=0,$ $Q={\cal O}(\lam^{1/2}),$ and $C={\cal O}(\lam),$
and that the sign of $Q$ depends on whether $F(\laminv)$ is
greater or less than $\pi/2.$ As $\lam$ increases further, ``A'' and
``B'' exchange relative positions every half period in the
sinusoidal oscillations of Eq.~\FIIUV. Numerical evidence for the global
minimum solution ``C,'' nearer the Skyrmion, is presented in the text.

\bye